\documentclass[12pt,a4paper]{article}
\usepackage[pdftex]{graphicx,xcolor}
\usepackage{amsmath,amssymb,graphicx,array}
\usepackage{setspace}
\usepackage[numbers]{natbib}

\RequirePackage{lineno}

\usepackage{footnote}

\newcommand{\ignore}[1]{}

\renewcommand{\title}{Hydro-mechanical network modelling of particulate composites}

\begin{document}

\begin{center} \begin{LARGE} \textbf{\title} \end{LARGE} \end{center}

\begin{center}

  Ignatios Athanasiadis, Simon J. Wheeler and Peter Grassl$^*$

  School of Engineering, University of Glasgow, Glasgow, UK

  $^*$Corresponding author: Email: peter.grassl@glasgow.ac.uk, Phone: +44 141 330 5208

  
\end{center}


\section*{Abstract}
Differential shrinkage in particulate quasi-brittle materials causes microcracking which reduces durability in these materials by increasing their mass transport properties.
A hydro-mechanical three-dimensional periodic network approach was used to investigate the influence of particle and specimen size on the specimen permeability.
The particulate quasi-brittle materials studied here consist of stiff elastic particles, and a softer matrix  and interfacial transition zones between matrix and particles exhibiting nonlinear material responses.
An incrementally applied uniform eigenstrain, along with a damage-plasticity constitutive model, are used to describe the shrinkage and cracking processes of the matrix and interfacial transition zones.
The results showed that increasing particle diameter at constant volume fraction increases the crack widths and, therefore, permeability, which confirms previously obtained 2D modelling results.
Furthermore, it was demonstrated that specimen thickness has, in comparison to the influence of particle size, a small influence on permeability increase due to microcracking.

\section{Introduction}
Microcracking due to particle restrained shrinkage significantly increases the permeability of porous quasi-brittle materials, which often reduces the durability of these materials. 
For instance for cementitious composites, microcracking due to particle (aggregate) restrained shrinkage has been experimentally observed in \cite{BisMie02,WonZobBue09,MarSas14,WuWonBue15,MarSasLin16} and has shown to increase mass transport properties such as permeability and sorptivity \cite{WonZobBue09,WuWonBue15}.
Numerically, the initiation of microcracking due to particle restrained shrinkage was studied in \cite{GraWonBue10,LagJouDes11,IdiBisCab12}.
In some of these studies, it was shown that the width of cracks produced by particle restrained shrinkage depends strongly on the size of particles \cite{GraWonBue10,IdiBisCab12}.
In the two-dimensional numerical modelling, particles were often idealised as cylindrical particles and all cracks were assumed to penetrate completely the specimen thickness.
Therefore, in these two-dimensional analyses, increase of crack width resulted in a significant increase in permeability.

In experiments of irregular particulate composites such as concrete, permeability was measured by applying a unidirectional pressure gradient of either water or gas across the specimen thickness.
In these thicker specimens, it can be assumed that complicated 3D fracture networks are generated due to differential shrinkage.
It appears to be reasonable to expect that part of these networks are not connecting the opposite sides of the specimen and, therefore, do not equally contribute to the increase of permeability.
Even for crack paths connecting opposite sides of the specimen, the crack widths along the path might vary.
Therefore, it can be expected that the thicker the specimen is, the smaller the increase of permeability due to cracking will be.
For instance, recent experimental studies in \cite{WuWonBue15} for nonuniform drying shrinkage showed that permeability depends on specimen thickness.
This dependence on specimen thickness could be due to the nonuniformity of the shrinkage strain, resulting in thickness dependent patterns of microcracking in the specimen or due to the variation of crack openings along random crack planes.

The aim of the present study was to investigate numerically the separate influences of specimen thickness and particle size on the increase of permeability due to particle restrained shrinkage induced microcracking.
For this purpose, a new coupled hydro-mechanical periodic network approach consisting of coupled structural and transport networks was developed.
 Periodic cells are known in the area of homogenisation \cite{MieKoc02,KanForGal03}, where it has been shown that periodic boundaries result in faster convergence of properties with increasing cell size than boundaries subjected to displacement or traction conditions.
One of the new features of the present periodic approach is that not only the periodic displacement/pressure-gradient conditions were applied, but also the three-dimensional network structure has been chosen to be periodic for both the structural and the transport networks. Peviously, combinations of periodic network structure and periodic displacement conditions were developed only for two-dimensional structural networks \citep{GraJir10}.
The new three-dimensional coupled periodic approach allows for describing fracture patterns and resulting increases in conductivity independent of boundaries.
For the transport network, the constitutive models were based on Darcy's law combined with a cubic law to model the influence of cracking on permeability \cite{GraBol16}.

\section{Method}\label{sec:method}
The present numerical method for investigating the influence of particle restrained shrinkage induced microcracking on transport properties was based on three-dimensional periodic hydro-mechanical networks of structural and transport elements.
For the constitutive model of structural elements, a damage-plasticity model was used \cite{GraDav11}.
The transport constitutive model used was Darcy's law combined with a cubic law \cite{WitWanIwaGal80} to model the increase of permeability due to fracture.
The new feature of the present method is the extension of the coupled network approach proposed in \cite{GraBol16} to a periodic cell of the shape of a rectangular cuboid, which uses periodic network structures and periodicity requirements for the displacements, rotations and pressures.
In addition to nodal degrees of freedom in the form of displacements, rotations and fluid pressures of nodes inside the cell, average strain and pressure gradient components are used to solve for unknown degrees of freedom of nodes outside the cell.
The use of a periodic cell with periodic network structure and periodicity requirements for the degrees of freedom has the advantage that for the mechanical network, crack patterns are independent of the cell boundaries, which is not the case for boundary value problems with either displacements or tractions applied to the boundaries.
The formulation of the hydro-mechanical periodic cell approach is conceptually based on work reported in \cite{GraJir10}.
However, the method in \cite{GraJir10} was limited to a two-dimensional structural network. Here, a three-dimensional coupled hydro-mechanical periodic network approach was proposed.   

\subsection{Discretisation}
\begin{figure}[h!]
\begin{center}
\includegraphics[height=6.cm]{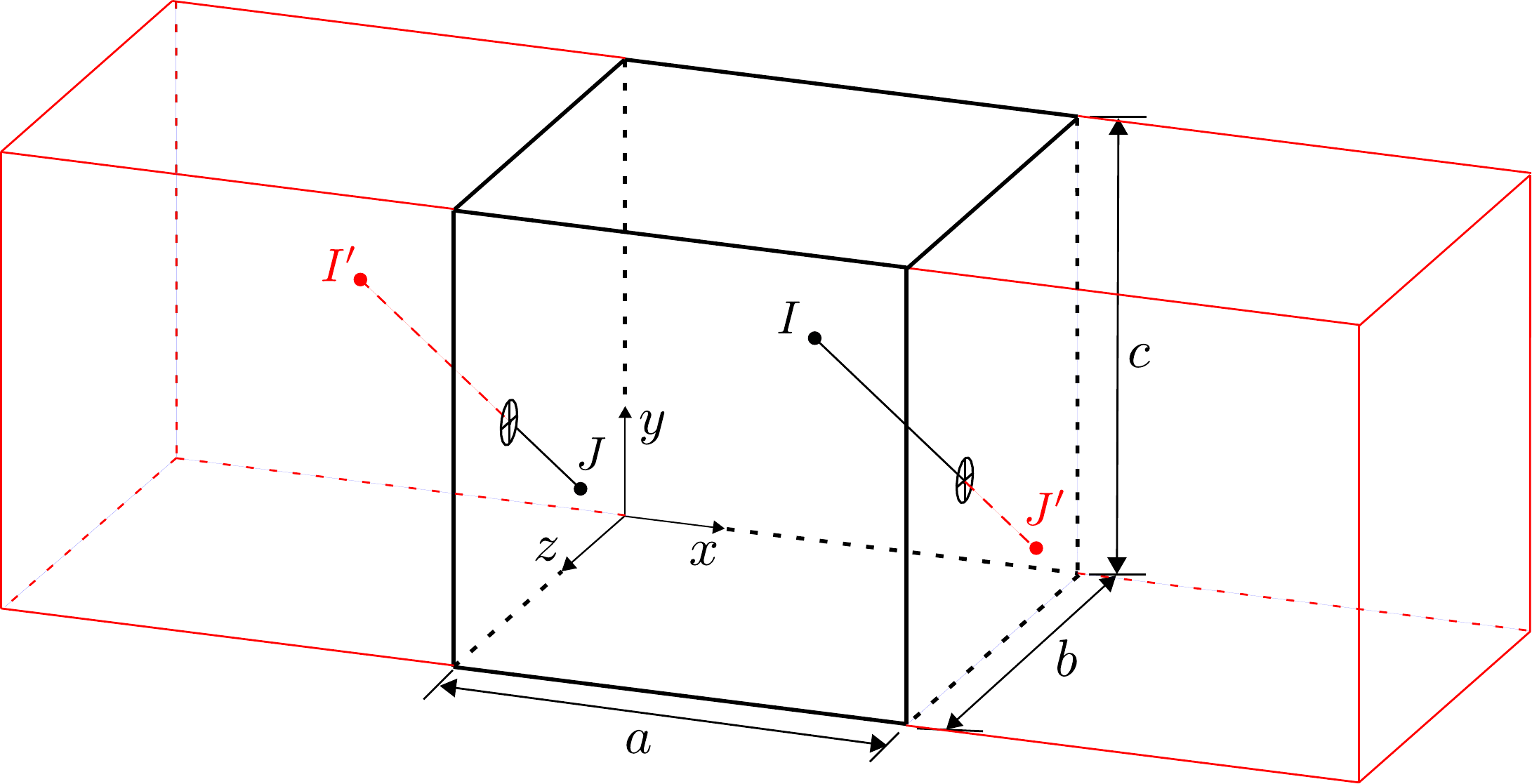}
\caption{3D periodic cell: main cell (thick lines) and 2 out of 26 neighbouring cells (thin lines) with two points $I$ and $J$ and their periodic images $I'$ and $J'$.}
\label{fig:3dBox}
\end{center}
\end{figure}

The periodic dual network approach was based on Delaunay and Voronoi tessellations of a set of points placed randomly within a rectangular cuboid shown in Figure~\ref{fig:3dBox} with thick lines.
The points were placed sequentially while enforcing a minimum distance $d_{\rm{min}}$ between all placed points.
Trial points that fail the minimum distance criterion were rejected.
The placement was terminated once the number of trials for placing one point exceeds the limit $N_{\rm{iter}}$.

Once the placement of points was completed, 26 periodic image points were generated for each successfully placed point within the cell using the translation rule
\begin{equation}\label{eq:pointPeriodic}
\mathbf{x}' = \mathbf{M} \mathbf{x} 
\end{equation}
where $\mathbf{x}$ and $\mathbf{x}'$ are the coordinate vector of the original point and one of the image points, respectively. Furthermore, $\mathbf{M}$ is the translation matrix defined as $\mathbf{M} = {\rm {diag}} [1 + k_{\rm x} a, 1 + k_{\rm y} b, 1 + k_{\rm z} c]$ where $k_{\rm x}, k_{\rm y}, k_{\rm z} \in \{-1,0,1\}$.
The $k_{\rm x}, k_{\rm y}$ and $k_{\rm z}$ coefficients define the direction of the shift from the original point to the image point.
The coordinates of the 26 image points are the result of the coordinate translations in (\ref{eq:pointPeriodic}) for all $k_{\rm x}, k_{\rm y}, k_{\rm z}$ combinations except for the case where $k_{\rm x} = k_{\rm y} = k_{\rm z} = 0$.
In Figure~\ref{fig:3dBox}, the cell with two of its 26 neighbours is shown.
The points $I$ and $J$ are examples of two randomly placed points satisfying the minimum distance requirements.
Points $I'$ and $J'$ are one of 26 sets of periodic image points of $I$ and $J$, respectively, whereby $I'$ was generated by a translation with $k_{\rm x} = -1$ and $k_{\rm y} = k_{\rm z} = 0$, and $J'$ with $k_{\rm x} = 1$ and $k_{\rm y} = k_{\rm z} = 0$.

All points within the cell and all periodic image points are used for the Delaunay and Voronoi tessellations. 
The Delaunay tessellation decomposes the domain into tetrahedra whose vertices coincide with the randomly placed points.
The Voronoi tessellation divides the domain into polyhedra associated with the random points~\cite{OkaBooSug00}.
Each polyhedron is the subset of the domain in which points are closer to the placed point that is associated with the polyhedron than all the other placed points.
Facets of Voronoi polyhedra form subsets of the 3D space, in which every location is equidistant from a pair of placed points and nearer to these two points than to any other point.
The edges of Delaunay tetrahedra connect pairs of placed points of Voronoi polyhedra with common facets.

\begin{figure}
  \begin{center}
    \begin{tabular}{ccc}
      \includegraphics[width=5.5cm]{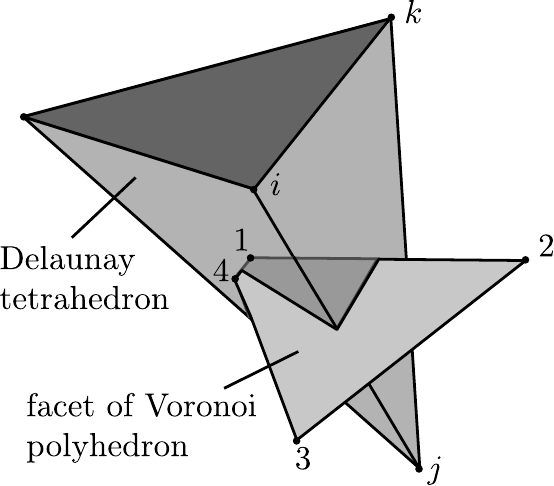} & \includegraphics[width=5cm]{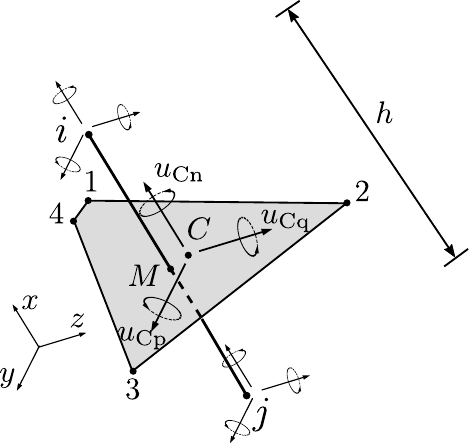} & \includegraphics[width=3cm]{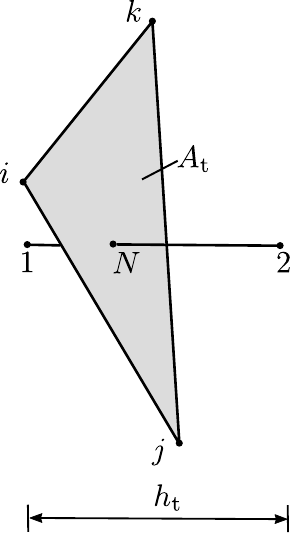}\\
      (a) & (b) & (c)
    \end{tabular}
\end{center}
    \caption{Discretisation: Spatial arrangement of structural and transport elements of the 3D transport-structural network approach showing a) geometrical relationship between Delaunay and Voronoi tessellations, (b) structural element with cross-section defined by the associated Voronoi facet and (c) transport element with cross-section defined by the associated Delaunay facet.}
\label{fig:3DTess}
\end{figure}

Delaunay and Voronoi tessellations were used to define the structural and transport elements \cite{GraBol16}.
In Figure~\ref{fig:3DTess}a, a Delaunay tetrahedron and the Voronoi facet associated with Delaunay edge $i$-$j$ are shown.
The structural elements were placed on the Delaunay edges with their mid-cross-sections defined by the facets of the Voronoi polyhedra (Figure~\ref{fig:3DTess}b).
Analogous to the structural network, the transport elements were placed on the edges of the Voronoi polyhedra, with their cross-sections formed by the facets of the Delaunay tetrahedra (Figure~\ref{fig:3DTess}(c)).

Because of the periodic image points, the tessellated space is larger than the main cell.
Therefore, edges of Delaunay tetrahedra and Voronoi polyhedra cross the cell boundaries.
In Figure~\ref{fig:3dBox}, an example of two elements crossing the cell boundaries is shown.
Here, the intersections of the elements $I$-$J'$ and $I'$-$J$ with the boundaries of the periodic cell are presented with a circle and a cross in the plane of the boundary.
The parts of the elements that lie outside the cell boundaries are shown with dashed lines and those inside by solid lines.
In the present periodic cell approach, all degrees of freedom of nodes inside the cell boundaries were involved in the system of equations that was used for determining the unknown degrees of freedom.
For elements crossing the boundary, the degrees of freedom of the nodes outside the cell were determined from those of the periodic image of the node inside the cell and additional information in the form of average strain and pressure gradients for the structural and transport problem, respectively. 

\begin{figure}[h!]
\begin{center}
\begin{tabular}{cc}
\includegraphics[height=5.cm]{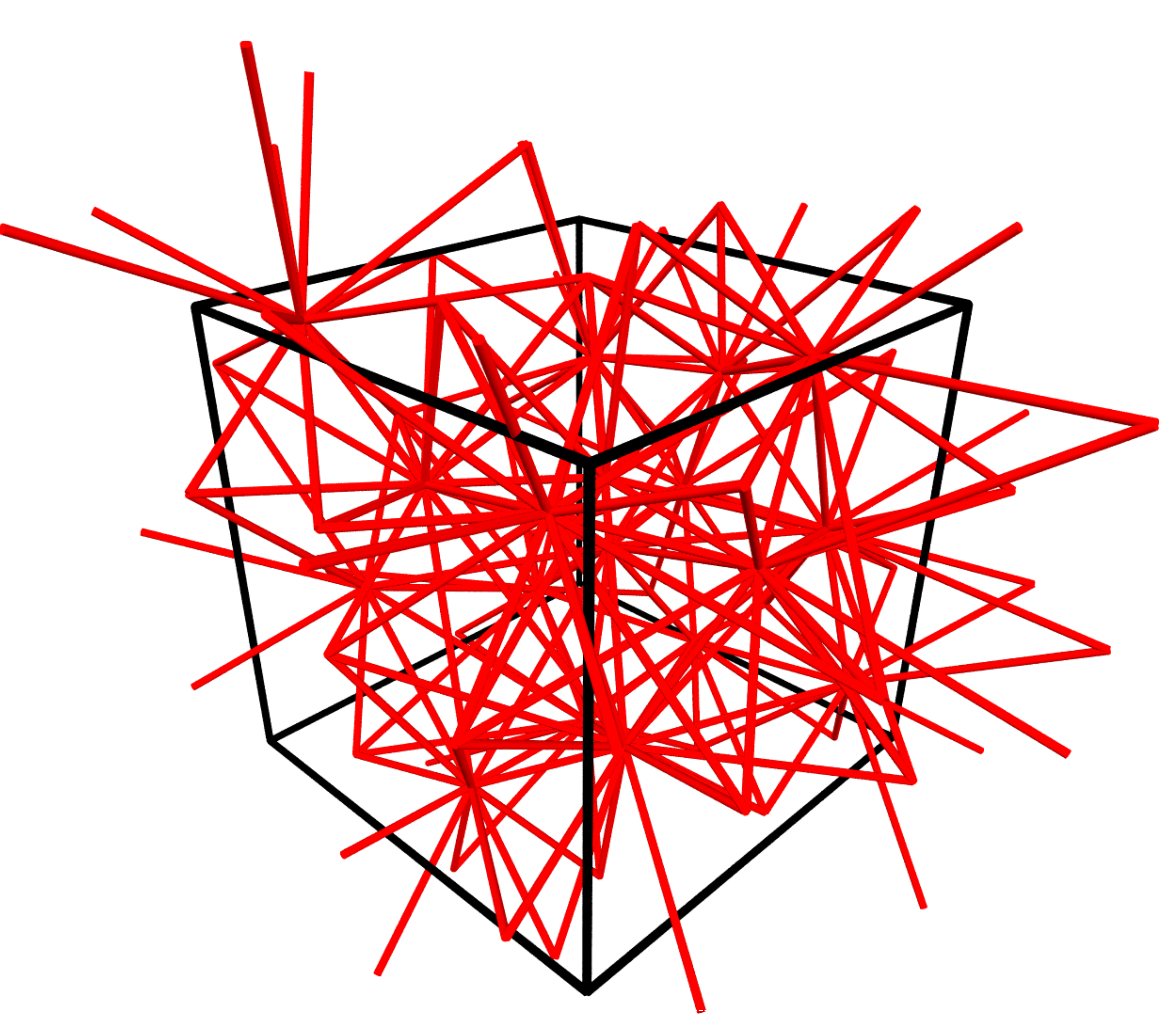}  & \includegraphics[height=5.cm]{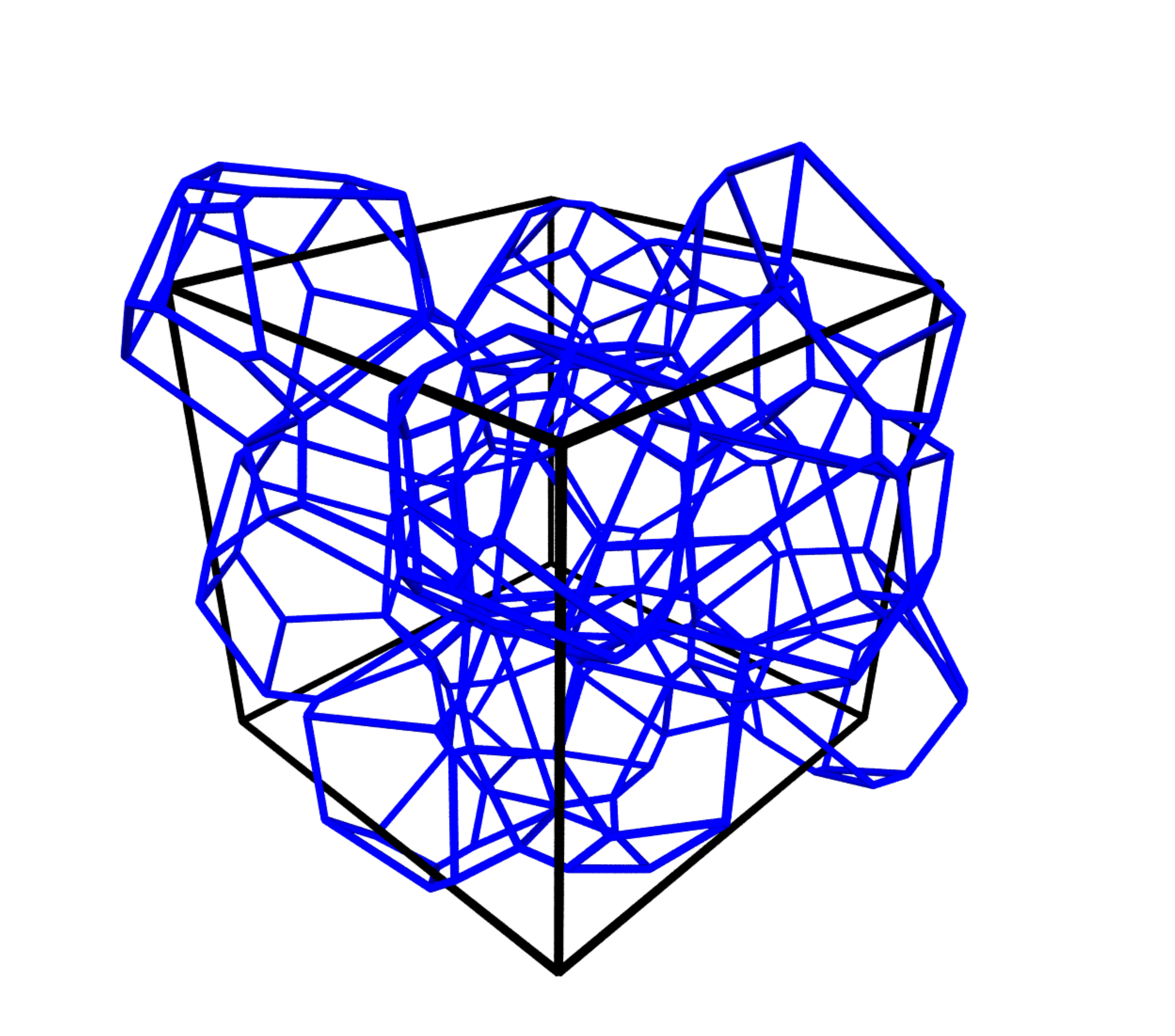}\\
(a) & (b) 
\end{tabular}
\caption{3D structural model including elements crossing the boundaries: (a) structural network and (b) edges of associated facets.}
\label{fig:crossing3DMech}
\end{center}
\end{figure}

\begin{figure}[h!]
\begin{center}
\begin{tabular}{cc}
\includegraphics[height=5.cm]{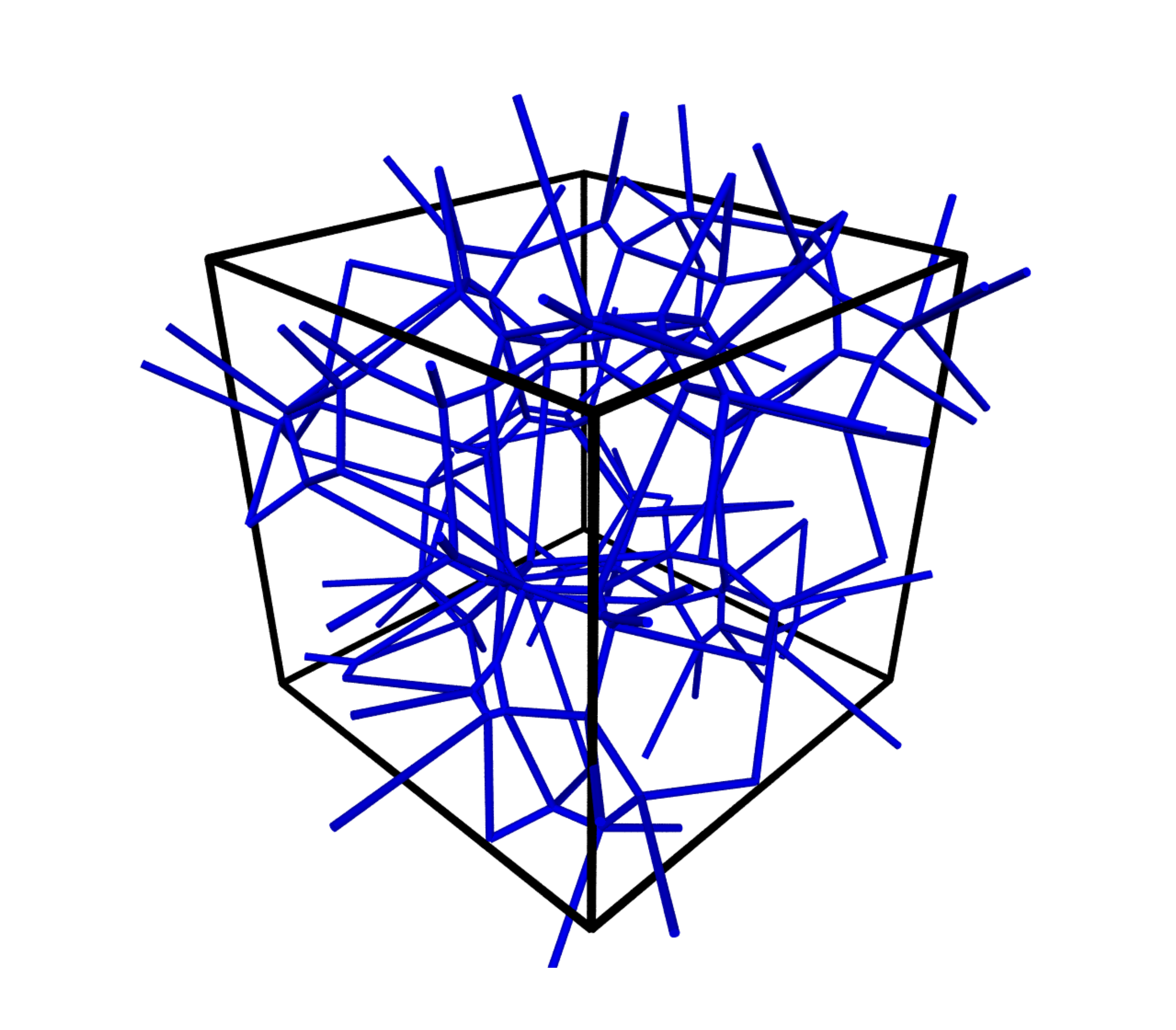} & \includegraphics[height=5.cm]{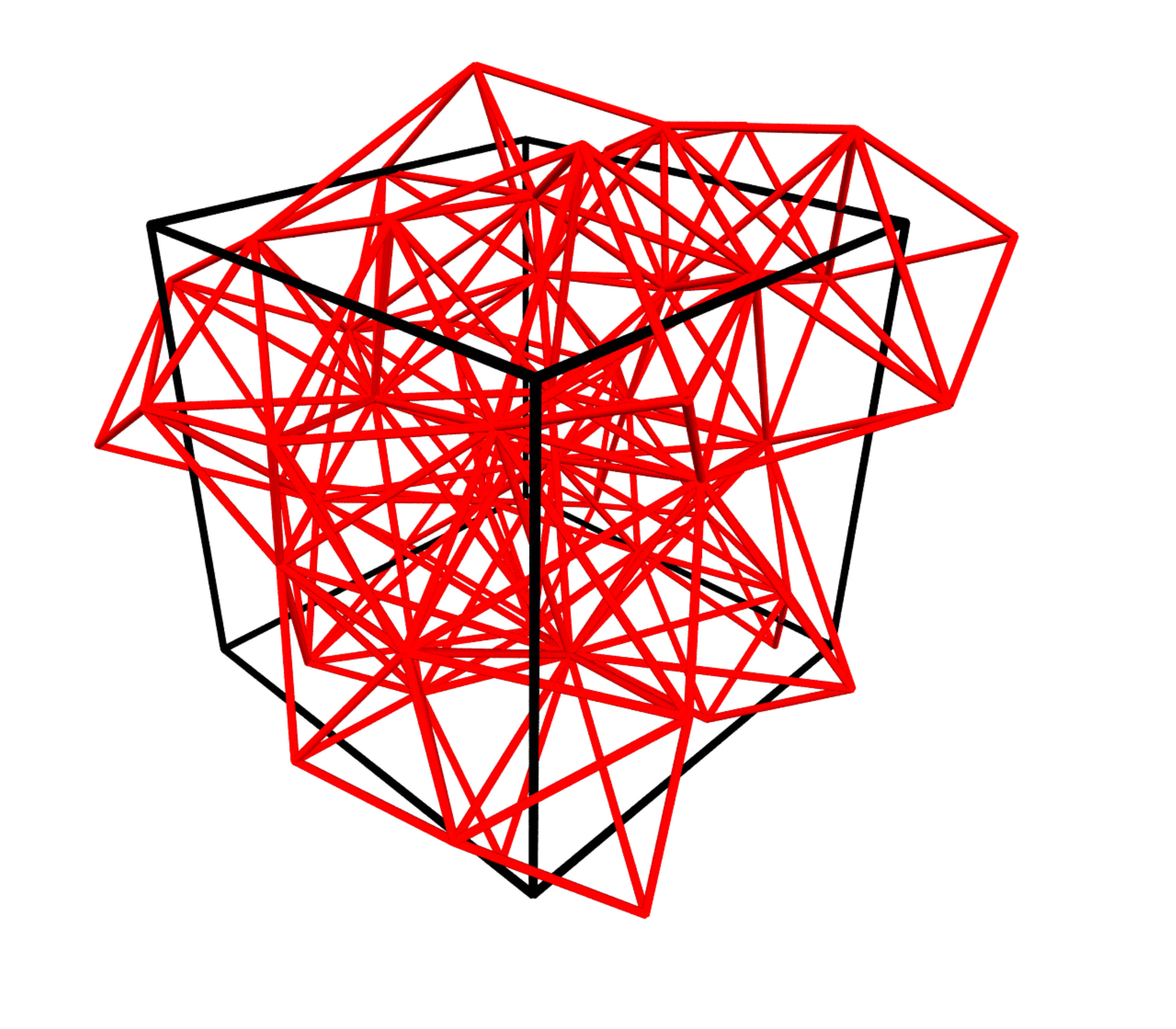}\\
 (a) & (b) 
\end{tabular}
\caption{3D transport model including elements crossing the boundaries: (a) transport network, (b) edges of associated facets.}
\label{fig:crossing3DTrans}
\end{center}
\end{figure}

Examples of structural and transport networks generated from the same set of random points are presented in Figures~\ref{fig:crossing3DMech}~and~\ref{fig:crossing3DTrans}, respectively.
For the structural network, the structural elements are shown in Figure~\ref{fig:crossing3DMech}a and the edges of the mid-cross-sections of the structural elements in Figure~\ref{fig:crossing3DMech}b.
The transport elements of the transport network are presented in Figure~\ref{fig:crossing3DTrans}a and the edges of the transport mid-cross-sections are shown in Figure~{\ref{fig:crossing3DTrans}}b.
From these examples of structural and transport networks, several interesting features of this coupled periodic cell are visible.
Firstly, both structural and transport elements (Figures~\ref{fig:crossing3DMech}a~and~\ref{fig:crossing3DTrans}a) are either located inside the cell or cross its boundaries.
No elements are entirely placed outside the periodic cell. 
The edges of the cross-sections of structural and transport elements are either inside the cell, cross the boundary or lie entirely outside the periodic cell.
Therefore, the network of structural elements in Figure~\ref{fig:crossing3DMech}a is smaller than the network of edges of the cross-sections of transport elements shown in Figure~\ref{fig:crossing3DTrans}b. Analogue to the structural network, the network of transport elements in Figure~\ref{fig:crossing3DTrans}a is smaller than the network of the edges of cross-sections of the structural network in Figure~\ref{fig:crossing3DMech}b.
For the coupling of the two networks, cross-section edges which are entirely outside the cell require special consideration. In the present network approach, a one way coupling approach was used in which crack openings obtained from the structural network were used to compute the conductivities of transport elements.
Therefore, for the transport network, crack openings associated with cross-sectional edges outside the cell were assumed to be equal to those of the corresponding cross-sectional edges located inside the cell \cite{Ath17}.
Details regarding how the conductivity was calculated are presented in section~\ref{sec:transMat}. The only input parameters required for the discretisation of the periodic cell are the minimum distance $d_{\rm{min}}$ and the maximum number of trials to place one point $N_{\rm iter}$. These parameters control the average lengths of structural and transport elements.
The greater $N_{\rm iter}$ is, the smaller is the standard deviation of the element lengths up to the stage at which the domain is almost saturated with points and and increase of $N_{\rm iter}$ will result in small changes of the number of placed points.

\subsection{Structural network}
The three-dimensional structural network was designed to approximate the quasi-static equilibrium equation without body force \cite{Strang86}, which is 
\begin{equation} \label{eq:stressEqui}
\nabla \boldsymbol{\sigma}^{\rm c} = \boldsymbol{0}
\end{equation}
where $\nabla$ is the divergence operator and $\boldsymbol{\sigma}^{\rm c}$ is the continuum stress.

\subsubsection{Structural element} \label{sec:structuralElement}
The structural element formulation for elements which lie entirely in the periodic cell is identical to the one presented in \cite{GraBol16}.
However, for elements crossing the boundary of the periodic cell, a new element formulation was introduced.
To be able to explain this new feature, the standard formulation is presented first.
The discrete version of (\ref{eq:stressEqui}) for the structural element shown in Figure~\ref{fig:3DTess}(b) is
\begin{equation}
\mathbf{K} \mathbf{u}_{\rm e} = \mathbf{f}_{\rm s}
\end{equation}
where $\mathbf{K}$ is the stiffness matrix, $\mathbf{u}_{\rm e}$ are the vector of degrees of freedom and $\mathbf{f}_{\rm s}$ are the acting forces.
The formulation of the structural element is presented in the local coordinate system, i.e. the coordinate system ($x$, $y$ and $z$) of the nodal degrees of freedom coincides with the coordinate system ($n$, $p$ and $q$) of the quantities used for evaluating the constitutive response.
Each node has 3 translational ($u_{\rm x}$, $u_{\rm y}$ and $u_{\rm z}$) and 3 rotational ($\phi_{\rm x}$, $\phi_{\rm y}$ and $\phi_{\rm z}$) degrees of freedom.
The degrees of freedom of a structural element with nodes $i$ and $j$ are grouped in translational and rotational parts as $\mathbf{u}_{\rm e} = \left\{\mathbf{u}_{\rm t}^T, \mathbf{u}_{\rm r}^T \right\}^T$, where $\mathbf{u}_{\rm t} = \left\{\mathbf{u}_{\rm ti}^T, \mathbf{u}_{\rm tj}^T\right\}^T = \left\{ u_{\rm xi}, u_{\rm yi}, u_{\rm zi}, u_{\rm xj}, u_{\rm yj}, u_{\rm zj} \right\}^T$ and $\mathbf{u}_{\rm r} = \left\{\mathbf{u}_{\rm ri}^T, \mathbf{u}_{\rm rj}^T\right\}^T = \left\{ \phi_{\rm xi}, \phi_{\rm yi}, \phi_{\rm zi}, \phi_{\rm xj}, \phi_{\rm yj}, \phi_{\rm zj} \right\}^T$. 
These degrees of freedom $\mathbf{u}_{\rm t}$ and $\mathbf{u}_{\rm r}$ are used to determine displacement discontinuities $\mathbf{u}_{\rm C} = \left\{u_{\rm Cn}, u_{\rm Cp}, u_{\rm Cq}\right \}^T$ at point $C$ by rigid body kinematics \cite{Kaw78} as 
\begin{equation}\label{eq:jump}
\mathbf{u}_{\rm C} = \mathbf{B} \mathbf{u}_{\rm e} = \mathbf{B}_{\rm 1} \mathbf{u}_{\rm t} + \mathbf{B}_{\rm 2} \mathbf{u}_{\rm r}
\end{equation}
where $\mathbf{B} = \left\{\mathbf{B}_1, \mathbf{B}_2\right\}^T$, $\mathbf{B}_{1}$ and $\mathbf{B}_{2}$ are two matrices containing the rigid body information for the nodal translations and rotations, respectively, which are
\begin{equation}\label{eq:BTrans}  
\mathbf{B}_{\rm 1} =
\begin{pmatrix}
  -\mathbf{I} & \mathbf{I}
\end{pmatrix}
\end{equation}
and
\begin{equation}\label{eq:BRot}  
\mathbf{B}_{\rm 2} =
\begin{pmatrix}
  0 & -e_{\rm q} & e_{\rm p} & 0 & e_{\rm q} & -e_{\rm p}\\
  e_{\rm q} & 0 & -h/2 & -e_{\rm q} & 0 & -h/2\\
  -e_{\rm p} & h/2 & 0 & e_{\rm p} & h/2 & 0
\end{pmatrix}
\end{equation}
Here, $\mathbf{I}$ is a $3\times3$ unity matrix.
In (\ref{eq:BRot}), $e_{\rm p}$ and $e_{\rm q}$ are the eccentricities between the midpoint of the network element and the centroid $C$ in the directions $p$ and $q$ of the local coordinate system, respectively (Figure~\ref{fig:3DTess}b).
The local coordinate system is defined by the direction $n$, which is parallel to the axis of the element, and $p$ and $q$, which are chosen as the two principal axes of the mid-cross-section. 

The displacement jump $\mathbf{u}_{\rm C}$ in (\ref{eq:jump}) is transformed into strains $\boldsymbol{\varepsilon} = \left\{\varepsilon_{\rm n}, \varepsilon_{\rm p}, \varepsilon_{\rm q} \right\}^T = \mathbf{u}_{\rm C}/h$, where $h$ is the length of the structural element.
The strains are related to stresses $\boldsymbol{\sigma} = \left\{\sigma_{\rm n}, \sigma_{\rm p}, \sigma_{\rm q} \right\}^T$ by means of a material stiffness $\mathbf{D} = \left(1-\omega\right) \mathbf{D}_{\rm e}$, where $\mathbf{D}_{\rm e} = \mathrm{diag} \left\{E, E, E \right \}$. Here, $E$ is the Young's modulus and $\omega$ is the damage variable, which is further discussed in Section~\ref{sec:structuralMaterial}.
For the present elastic stiffness matrix $D_{\rm e}$, Poisson's ratio equal to zero is obtained and the structural network is elastically homogeneous under uniform modes of straining.

For the case that the global coordinate system coincides with the local one, the element stiffness matrix is
\begin{equation}
  \mathbf{K} = \dfrac{A}{h}
  \begin{pmatrix}
    \mathbf{B}_{\rm 1}^{\rm T} \mathbf{D} \mathbf{B}_{\rm 1} &  \mathbf{B}_{\rm 1}^{\rm T} \mathbf{D} \mathbf{B}_{\rm 2}\\
    \mathbf{B}_{\rm 2}^{\rm T} \mathbf{D} \mathbf{B}_{\rm 1} &  \mathbf{B}_{\rm 2}^{\rm T} \mathbf{D} \mathbf{B}_{\rm 2}
  \end{pmatrix}
  +
  \begin{pmatrix}
    \mathbf{0} & \mathbf{0}\\
    \mathbf{0} & \mathbf{B}_{\rm 1}^{\rm T} \mathbf{K_{\rm r}} \mathbf{B}_{\rm 1}         
  \end{pmatrix}
 \end{equation}
Here, $\mathbf{K}_{\rm r}$ is a matrix containing the rotational stiffness at point $C$ defined as
\begin{equation}\label{eq:Kr}
  \mathbf{K}_{\rm r} = \dfrac{(1-\omega)E}{h}
  \begin{pmatrix}
    I_{\rm p} & 0 & 0\\
    0 & I_{1} & 0\\
    0 & 0 & I_{2}
  \end{pmatrix}  
\end{equation}
Here, $I_{\rm p}$ is the polar moment of area, and $I_{1}$ and $I_{2}$ are the two principal second moments of area of the cross-section.
The factor $(1-\omega)$ in (\ref{eq:Kr}) ensures that the rotational stiffness reduces to zero for a fully damaged cross-section ($\omega=1$).
The stiffness matrix is then expressed in the global coordinate system by means of rotation matrices as described for instance in \cite{McgGalZie00}.

The above element formulation for structural elements entirely located in the periodic cell is identical to the one described in \cite{GraBol16}.
For elements crossing the cell boundaries, a special formulation is required.
For these elements, the degrees of freedom of the nodes outside the cell are determined from the degrees of freedom of the periodic image inside the cell and the average strain $\mathbf{E} = \left\{E_{\rm x}, E_{\rm y}, E_{\rm z}, E_{\rm yz}, E_{\rm zx}, E_{\rm yx}\right\}^{T}$. 
Here, $E_{\rm x}$, $E_{\rm y}$ and $E_{\rm z}$ are the average normal strains in the $x$, $y$ and $z$ direction,
respectively and $E_{\rm {yz}}$, $E_{\rm {zx}}$, $E_{\rm {yx}}$ are the average engineering shear strain components.
For an illustration of the coordinate system $x$, $y$ and $z$, see Figure~\ref{fig:3dBox}.
The translations of a node outside the cell is
\begin{equation}
u'_{\rm x} = u_{\rm x} + ak_{\rm x}E_{\rm x} + ck_{\rm z}E_{\rm {zx}} + bk_{\rm y}E_{\rm {yx}}
\label{eq:dispUTrans}
\end{equation}
\begin{equation}
u'_{\rm y} = u_{\rm y} + bk_{\rm y}E_{\rm y} + ck_{\rm z}E_{\rm {yz}}
\label{eq:dispVTrans}
\end{equation}
\begin{equation}
u'_{\rm z} = u_{\rm z} + ck_{\rm z}E_{\rm z} 
\label{eq:dispWTrans}
\end{equation}
where the translation presented without and with the prime symbol are those of the nodes located within and outside the cell, respectively.
Note that the contributions of the average shear strains $E_{\rm {zx}}$ and $E_{\rm {yx}}$ have been included only in the displacements in the $x$ direction and the contribution of $E_{\rm {yz}}$ has been included only in the displacement in the $y$-direction.
This is justified because rigid body rotations of the entire cell are arbitrary.
One node of the network is fully fixed in order to prevent rigid body rotation and translation.

Consider the element $IJ'$ in Figure~\ref{fig:3dBox}. Node $J'$ is outside the cell and its periodic image $J$ is inside the cell.
Making use of (\ref{eq:dispUTrans}), (\ref{eq:dispVTrans}) and (\ref{eq:dispWTrans}) and assuming that $\phi_{xJ}$~=~$\phi_{xJ'}$, $\phi_{yJ}$~=~$\phi_{yJ'}$ and $\phi_{zJ}$~=~$\phi_{zJ'}$, the transformation rule giving the translations and rotations of the two ends $I$ and $J'$ of a structural element $IJ'$ crossing a cell boundary is
\begin{equation} \label{eq:transDofs}
\begin{pmatrix}
  \mathbf{u}_{I} \\
  \mathbf{u}_{J'} \\
  \mathbf{r}_{I} \\
  \mathbf{r}_{J'}
\end{pmatrix}
=
\mathbf{T}_{\rm m}
\begin{pmatrix}
  \mathbf{u}_{I}\\
  \mathbf{u}_{J}\\
  \mathbf{r}_{I}\\
  \mathbf{r}_{J}\\
  \mathbf{E}
\end{pmatrix}
\end{equation}
where $\mathbf{u}_{I}$, $\mathbf{r}_{I}$, $\mathbf{u}_{J}$, $\mathbf{r}_{J}$, $\mathbf{u}_{J'}$ and $\mathbf{r}_{J'}$ are the vectors containing translational and rotational degrees of freedom of nodes $I$, $J$ and $J'$, respectively. The node $J$ is the periodic image of point $J'$ inside the cell.
The transformation matrix $\mathbf{T}_{\rm m}$ is of size $12 \times 18$ and has the form
\begin{equation}\label{eq:tmStruc}
\mathbf{T}_{\rm m}
=
\begin{bmatrix}
  \mathbf{I} & \mathbf{0} & \mathbf{0} & \mathbf{0} & \mathbf{0} & \mathbf{0} \\
  \mathbf{0} & \mathbf{I} & \mathbf{0} & \mathbf{0} & \mathbf{k}_{\rm 21} & \mathbf{k}_{\rm 22} \\
  \mathbf{0} & \mathbf{0} & \mathbf{I} & \mathbf{0} & \mathbf{0} & \mathbf{0}\\
   \mathbf{0} & \mathbf{0} & \mathbf{0} & \mathbf{I} & \mathbf{0} & \mathbf{0}\\
\end{bmatrix}
\end{equation}
The sub-matrices $\mathbf{k}_{\rm 21}$ and $\mathbf{k}_{\rm 22}$ are $3 \times 3$ matrices which contain information about the transformation of the nodal translations due to the average strains.
They are defined as
\begin{equation}
\mathbf{k}_{21}
 =
\begin{bmatrix}
 ak_{\rm x} &  0 & 0\\ 
 0 & bk_{\rm y} &  0\\ 
0 & 0 & ck_{\rm z}\\ 
\end{bmatrix}
\label{eq23}
\end{equation}
and
\begin{equation}
\mathbf{k}_{22}
 =
\begin{bmatrix}
0 & ck_{\rm z} & bk_{\rm y}\\ 
ck_{\rm z} & 0 & 0\\ 
0 & 0 & 0\\ 
\end{bmatrix}
\label{eq24}
\end{equation}
If (\ref{eq:transDofs}) is combined with (\ref{eq:jump}) for calculating the displacement jump, the transformation matrix $\mathbf{T}_{\rm m}$ multiplies matrix $\mathbf{B}$ from the right.
It follows from duality that the internal forces must be multiplied by $\mathbf{T}_{\rm  m}^{\rm T}$ from the left, before the evaluation of the equilibrium conditions.
Hence, the original $ 12 \times 12$ stiffness matrix $\mathbf{K}$ of the non-periodic element is now transformed into the $18 \times 18$ matrix $\mathbf{T}_{\rm m}^{\rm T} \mathbf{K} \mathbf{T}_{\rm m}$.

With the present approach, average stresses and strains are prescribed by means of the additional six average strain components introduced in the formulation of the periodic cell.
Finally, the total number of degrees of freedom are six times the number of nodes positioned within the cell boundaries plus six additional degrees of freedom, which correspond to the six average strain components.

\subsubsection{Structural material}\label{sec:structuralMaterial}
The constitutive model for the structural material is based on a damage-plasticity framework \cite{GraDav11}, which is capable of reproducing the important features of the response of quasibrittle materials in tension and compression.
The strains are related to the nominal stress $\boldsymbol{\sigma} = \left\{\sigma_{\rm n}, \sigma_{\rm p}, \sigma_{\rm q} \right\}^T$ as 
\begin{equation}\label{eq:totStress}
\boldsymbol{\sigma} = \left(1-\omega\right) \mathbf{D}_{\rm e} \left(\boldsymbol{\varepsilon}-\boldsymbol{\varepsilon}_{\rm p} - \boldsymbol{\varepsilon}_{\rm s}\right) = \left(1-\omega\right) \bar{\boldsymbol{\sigma}}
\end{equation}
where $\omega$ is the damage variable, $\mathbf{D}_{\rm e}$ is the elastic stiffness, $\boldsymbol{\varepsilon}_{\rm p} = \left\{\varepsilon^{\rm p}_{\rm n}, \varepsilon^{\rm p}_{\rm p}, \varepsilon^{\rm p}_{\rm q} \right\}^T$ is the plastic strain and $\bar{\boldsymbol{\sigma}}$ is the effective stress.
Furthermore, $\boldsymbol{\varepsilon}_{\rm s} = \left\{\varepsilon_{\rm s}, 0, 0\right\}^T$ is the shrinkage strain which was used in this study to initiate microcracking.

\begin{figure}
\begin{center}
  \includegraphics[width=10cm]{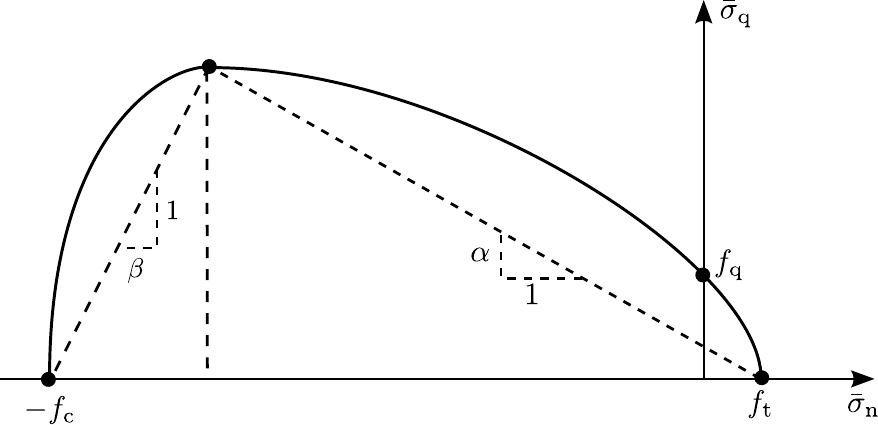}
\end{center}
\caption{Yield surface for $q = 1$. The surface is controlled by the tensile strength $f_{\rm t}$, the compressive strength $f_{\rm c}$ and two parameters $\alpha$ and $\beta$ which determine the out-of-roundness of the two ellipses which form the surface.}
\label{fig:yieldSurf}
\end{figure}

The plasticity model used to determine the effective stress is independent of damage.
The model is described by the yield function (\ref{eq:yieldGeneral}), flow rule (\ref{eq:plastFlow}), evolution law for the hardening variable (\ref{eq:evolution}) and loading unloading conditions (\ref{eq:loadUn}):
\begin{equation} \label{eq:yieldGeneral}
f = F\left(\bar{\boldsymbol{\sigma}},\kappa \right)
\end{equation}
\begin{equation}\label{eq:plastFlow}
\dot{\boldsymbol{\varepsilon}}_{\rm p} = \dot{\lambda} \dfrac{\partial g} {\partial \bar{\boldsymbol{\sigma}}}
\end{equation}
\begin{equation}\label{eq:evolution}
\dot{\kappa} = \dot{\lambda} h_{\kappa}
\end{equation}
\begin{equation}\label{eq:loadUn}
f \leq 0 \mbox{,} \hspace{0.5cm} \dot{\lambda} \geq 0 \mbox{,} \hspace{0.5cm} \dot{\lambda} f = 0
\end{equation}
Here, $f$ is the yield function, $\kappa$ is the hardening variable, $g$ is the plastic potential, $h_{\kappa}$ is the evolution law for the hardening parameter and $\dot{\lambda}$ is the rate of the  plastic multiplier.
The yield function of the two stress variables $\bar{\sigma}_{n}$ and $\bar{\sigma}_{q} = \sqrt{\bar{\sigma}_{\rm s}^2 + \bar{\sigma}_{\rm t}^2}$ is 
\begin{equation}\label{eq:yield}
f = \left\{ \begin{array}{ll} \alpha^2\bar{\sigma}_{\rm n}^2 + 2 \dfrac{\alpha^2\left(f_{\rm c} - \alpha \beta f_{\rm t}\right)}{1+\alpha \beta} q \bar{\sigma}_{\rm n} + \bar{\sigma}_{\rm q}^2 - \dfrac{2 \alpha^2 f_{\rm c}f_{\rm t} + \alpha^2 \left(1-\alpha \beta \right) f_{\rm t}^2}{1+\alpha \beta} q^2 & \hspace{0.5cm} \mbox{if $\bar{\sigma}_{\rm n} \geq \dfrac{f_{\rm c} - \alpha \beta f_{\rm t}}{1+\alpha \beta} q$} \vspace{0.5cm}\\
\dfrac{\bar{\sigma}_{\rm n}^2}{\beta^2} + 2 \dfrac{f_{\rm c} - \alpha \beta f_{\rm t}}{1+\alpha \beta} q \bar{\sigma}_{\rm n} + \bar{\sigma}_{\rm q}^2 + \dfrac{\left(1-\alpha^2 \beta^2\right) f_{\rm c}^2 -2 \alpha \beta \left(1+\alpha \beta\right) f_{\rm c} f_{\rm t} }{\beta^2 \left(1+\alpha \beta \right)} q^2 & \hspace{0.5cm} \mbox{if $\bar{\sigma}_{\rm n} < \dfrac{f_{\rm c} - \alpha \beta f_{\rm t}}{1+\alpha \beta} q$} \end{array} \right.
\end{equation}
where $f_{\rm t}$ and $f_{\rm c}$ are the tensile and compressive strengths, respectively, and $\alpha$ and $\beta$ are the friction angles shown in Fig.~\ref{fig:yieldSurf} for $f = 0$ and $q=1$, which controls the hardening.
It is defined as
\begin{equation}
q = \exp\left(\dfrac{\kappa}{A_{\rm h}}\right)
\end{equation}
where $A_{\rm h}$ is an input parameter. For the onset of plastic flow $\kappa = 0$ and $q = 1$. 

The stress dependent parts of the plastic potential $g$ in the non-associated flow rule in (\ref{eq:plastFlow}) are the same as those of the yield surface $f$ except that $\alpha$ is replaced by $\psi$:
\begin{equation}\label{eq:plast}
g = \left\{ \begin{array}{ll} \psi^2 \bar{\sigma}_{\rm n}^2 + 2  \dfrac{\psi^2 \left(f_{\rm c} - \psi \beta f_{\rm t}\right)}{1+\psi \beta} q \bar{\sigma}_{\rm n} + \bar{\sigma}_{\rm q}^2 & \hspace{0.5cm} \mbox{if $\bar{\sigma}_{\rm n} \geq \dfrac{f_{\rm c} - \psi \beta f_{\rm t}}{1+\psi \beta} q$} \vspace{0.5cm}\\
\dfrac{\bar{\sigma}_{\rm n}^2}{\beta^2} + 2 \dfrac{f_{\rm c} - \psi \beta f_{\rm t}}{1+\psi \beta} q \bar{\sigma}_{\rm n} + \bar{\sigma}_{\rm q}^2 & \hspace{0.5cm} \mbox{if $\bar{\sigma}_{\rm n} < \dfrac{f_{\rm c} - \psi \beta f_{\rm t}}{1 + \psi \beta} q$} \end{array} \right.
\end{equation}
The smaller $\psi$ is, the smaller is the ratio of normal and shear components of plastic strains for $\bar{\sigma}_{\rm n} \geq \dfrac{f_{\rm c} - \psi \beta f_{\rm t}}{1+\psi \beta} q$.
The function $h_{\rm \kappa}$ in the evolution law in (\ref{eq:evolution}) is chosen as 
\begin{equation}
h_{\kappa} = \left |\dfrac{\partial g}{\partial \sigma_{\rm n}} \right |
\end{equation}
which is the absolute value of the normal component of the direction of the plastic flow.

The damage variable in (\ref{eq:totStress}) is determined by means of the damage history variable
\begin{equation} \label{eq:damageHistory}
\kappa_{\rm d} = \left \langle \varepsilon_{\rm pn} \right \rangle
\end{equation}
where $\langle . \rangle$ denotes the McAuley brackets (positive part of operator).
The function of the damage variable is derived from the stress-crack opening curve in pure tension ($\sigma_{\rm n} > 0$, $\sigma_{\rm q} = 0$).
For the damage-plasticity constitutive model, the vector of crack opening components is defined as
\begin{equation}
\mathbf{w}_{\rm c} = h \left( \boldsymbol{\varepsilon}_{\rm p} + \omega \left(\boldsymbol{\varepsilon}_{\rm n} - \boldsymbol{\varepsilon}_{\rm pn} \right)\right)
\end{equation}
For pure tension, the crack opening simplifies to
\begin{equation} \label{eq:crackOpening}
w_{\rm c} = h \left( \varepsilon_{\rm pn} + \omega \left(\varepsilon_{\rm n} - \varepsilon_{\rm pn} \right)\right)
\end{equation} 
where $h$ is the length of the network element (Figure~\ref{fig:3dBox}).
The stress-crack opening curve is
\begin{equation} \label{eq:softLaw}
\sigma_{\rm n} = f_{\rm t} \exp \left(-\dfrac{w_{\rm c}}{w_{\rm f}}\right)
\end{equation}
where $w_{\rm f}$ controls the initial slope of the exponential softening curve. It is related to the area under the stress-crack opening curve $G_{\rm F}$ as $w_{\rm f} = G_{\rm F}/f_{\rm t}$.
Setting (\ref{eq:softLaw}) equal to the first component of (\ref{eq:totStress}), a nonlinear equation of the damage $\omega$ is obtained, which is solved using the Newton-Raphson method.
For modelling the dependence of transport properties on cracking, permeability, which is part of the transport model described in section~\ref{sec:transportModel}, is made dependent on the absolute value of the crack opening  
\begin{equation} \label{eq:equivCrack}
\tilde{w}_{\rm c} = \left|w_{\rm c}\right|
\end{equation}
The structural constitutive model requires eight input parameters.
The Young's modulus of the lattice material $E$ controls the macroscopic Young's modulus.
The parameters of the plasticity part are $f_{\rm t}$, $f_{\rm c}$, $\alpha$, $\beta$, $\psi$ and $A_{\rm h}$.
Finally, $G_{\rm F}$ controls the amount of energy dissipated during cracking.

\subsection{Transport network}\label{sec:transportModel}
For the transport part of the model, a 3D network of 1D transport elements is used to discretise the stationary transport equation \cite{MaeIshKis08}
\begin{equation} \label{eq:flow}
\mathrm{div} \left(k \mathrm{grad} P_{\rm f} \right) = 0
\end{equation}
Here, $P_{\rm f}$ is the fluid pressure, $k$ is the conductivity. In (\ref{eq:flow}), gravitational effects are neglected.

\subsubsection{Transport elements}
Analogue to the structural network, different element formulations are used for elements which are located entirely in the cell and those that cross one of the faces of the cell.
For those inside the cell, the discrete form of (\ref{eq:flow}) for a 1D transport element shown in Figure~\ref{fig:3DTess}c is
\begin{equation}\label{eq:flowDiscrete}
\mathbf{k}_{\rm e} \boldsymbol{P}_{\rm{f}} = \mathbf{f}_{\rm e}
\end{equation}
where $\mathbf{k}_{\rm e}$ is the one-dimensional element conductivity and $\mathbf{f}_{\rm e}$ are the nodal flow rate vector \cite{LewMorTho96,BolBer04}. The degrees of freedom of the transport elements are the fluid pressures $\boldsymbol{P}_{\rm f} = \left\{P_{\rm f1}, P_{\rm f2}\right\}^T$.
Within the context of a one-dimensional finite element formulation \cite{LewMorTho96}, Galerkin's method is used to construct the elemental conductivity matrix as 
\begin{equation}\label{eq:conductivityMatrix}
\mathbf{k}_{\rm e} = k \dfrac{A_{\rm t}}{h_{\rm t}}
\begin{pmatrix}
1 & -1\\
-1 & 1
\end{pmatrix}
\end{equation}
Here, $A_{\rm t}$ is the  mid-cross-sectional area and $h_{\rm t}$ the length of the transport element shown in Figure~\ref{fig:3DTess}c.
\begin{figure}
  \begin{center}
    \includegraphics[width=4cm]{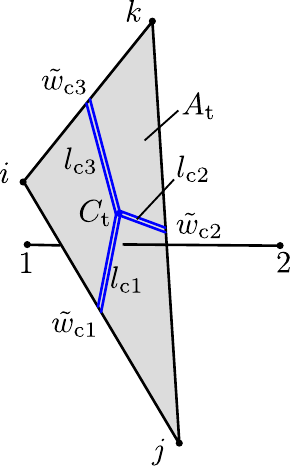}
\end{center}
  \caption{Influence of cracking on transport.}
\label{fig:transport3dCrack}
\end{figure}

For elements, which cross the faces of the cell, the element formulation is explained based on Figure~\ref{fig:3dBox} used earlier for the structural element. For instance, for the element $I'J$ in Figure~\ref{fig:3dBox}, the node $I'$ is outside the cell and $J$ is inside the cell.
The periodic image of $I'$ is $I$ which is located inside the cell, which is used together with average fluid pressure gradients to determine the fluid pressure of the node outside the cell.
The nodal fluid pressures of an element crossing the cell boundary are
\begin{equation} \label{eq25}
\begin{pmatrix}
{P}_{ \rm{f} I'} \\  {P}_{\rm{f}J}
\end{pmatrix}
=
\mathbf{T}_{\rm t}
\begin{pmatrix}
  P_{\rm{f}I}\\
  P_{\rm{f}J}\\
  \Delta P_{\rm \rm{f}x}/a\\
  \Delta P_{\rm \rm{f}y}/b \\
  \Delta P_{\rm \rm{f}z}/c
\end{pmatrix}
\end{equation}
where $\Delta P_{\rm{fx}}/a$, $\Delta P_{\rm{fy}}/b$  and $\Delta P_{\rm{fz}}/c$ are the average fluid pressure gradients along the $x$, $y$ and $z$ directions respectively.
Here, $a$, $b$ and $c$ are the dimensions of the periodic cell shown in Figure~\ref{fig:3dBox}.
Furthermore, $\mathbf{T}_{\rm t}$ is a transformation matrix of size $2 \times 5$ and has the form
\begin{equation} \label{eq26}
\mathbf{\rm T}_{\rm t} =
\begin{bmatrix}
 1  & 0 & ak_{\rm x} & bk_{\rm y} & ck_{\rm z}\\ 
 0 & 1 & 0 & 0 & 0\\ 
\end{bmatrix}
\end{equation}
For combining (\ref{eq26}) with (\ref{eq:flowDiscrete}), the transformation matrix $\mathbf{T}_{\rm t}$ multiplies matrix $\mathbf{k}_e$ from the right.
It follows from duality that the internal flux must be multiplied by $\mathbf{T}_{\rm t}^{\rm T}$ from the left, before the evaluation of the balance condition.
The conductivity matrix is evaluated as $\mathbf{T}_{\rm t}^{\rm T}\mathbf{k}_{\rm e} \mathbf{T}_{\rm t}$, where the original conductivity matrix $\mathbf{k}_{\rm e}$ is transformed from a $2 \times 2$ matrix to a $5 \times 5$ one.
The global conductivity matrix is assembled normally except for six rows and columns that relate the global degrees of freedom to its conjugate reaction flow rates.
As a result, the periodic cell can be subjected to arbitrary combinations of average flux or gradients.
The total number of unknown degrees of freedom is the total number of the nodes located in the interior of the periodic cell plus three global degrees of freedom controlling the average flux or pressure gradient in the three directions.

\subsubsection{Transport materials}\label{sec:transMat}
The conductivity matrix for the material of the transport elements is 
\begin{equation} \label{eq:conSplit}
k = k_0 + k_{\rm c}
\end{equation}
where $k_0$ is the initial conductivity of the undamaged material and $k_{\rm c}$ is the change of conductivity due to fracture.
The conducticity of the undamaged material is
\begin{equation}
k_0 = \dfrac{\rho \kappa_0}{\mu}
  \end{equation}
where $\rho$ is the density and $\mu$ is the dynamic viscosity of the fluid, and $\kappa_0$ is the intrinsic permeability.
In this work, the density and dynamic viscosity was set to $\rho=1000$~kg/m$^3$ and $\mu=0.001$~Pa~s, respectively, which corresponds to the values commonly used for water.

The second term $k_{\rm c}$ in (\ref{eq:conSplit}) models the increase of conductivity due to cracking using a cubic law based on the concept of flow through parallel plates \cite{WitWanIwaGal80} with a reduction factor $\xi$ for the presence of roughness of the wall surface \cite{AkhShaRaj12}.
A detailed description of the definition of $k_{\rm c}$ and its dependence on the crack openings of the structural network, which was used in this study, has already been presented in \cite{GraBol16}. However, since this is an important part of the model of the present study, it is shown here once more.
The term $k_{\rm c}$ is
\begin{equation}\label{eq:crackDiff}
k_{\rm c} = \xi \dfrac{\rho}{12 \mu A_{\rm t}} \sum_{i=1}^{3} \tilde{w}_{\rm{c}i}^3 l_{\rm{c}i}
\end{equation}
where $\tilde{w}_{\rm{c}i}$ and $l_{\rm{c}i}$ are the equivalent crack openings and crack lengths (see Figure~\ref{fig:transport3dCrack}) of neighbouring structural elements, which are located on the edges of the cross-section, and $\xi$ is a reduction factor which considers the reduction of flow for cracks with rough surfaces compared to that between smooth parallel plates.

Here, $\tilde{w}_{\rm c}$ is the magnitude of the crack opening $\mathbf{w}_{\rm c}$ defined in (\ref{eq:crackOpening}). 
The relation in (\ref{eq:crackDiff}) expresses the well known cubic law, which has shown to produce good results for transport in fractured geomaterials \cite{WitWanIwaGal80}. 
The way how crack openings in the structural elements influence the conductivity of a transport element is schematically shown in Figure~\ref{fig:transport3dCrack}.
For instance, for the transport element $o-p$, three structural elements ($i-k$, $k-j$ and $i-j$) bound the cross-section of the transport element.
Thus, the conductivity will be influenced by these three elements according to (\ref{eq:crackDiff}) in proportion to their equivalent crack widths and the crack lengths. This crack length (shown by blue double lines in Figure~\ref{fig:transport3dCrack}) is defined as the length from the midpoint of the structural element to the centroid $C_{\rm t}$ of the transport element cross-section.

\section{Analyses}
\subsection{Introduction}
The coupled structural transport network model described in section~\ref{sec:method} was applied to the analysis of particle restrained shrinkage of prisms made of a particulate quasi-brittle material consisting of particles, matrix and interfacial transition zones.
The matrix and interfacial transition zones are made of permeable quasi-brittle cohesive-frictional material, where the interfacial transition zone is weaker and more permeable than the matrix.
The particles are elastic, and stiffer and less permeable than the matrix.
The input parameters for the three material phases are presented in Table~\ref{tab:input}.
The properties of the different phases are mapped onto a coupled periodic network.
Network elements with both nodes within the same particle are given the property of particles. Those elements with nodes in different particles or one node in a particle and the other in the matrix are assigned the properties of the interfacial transition zone.
Here, it is assumed that the thickness of the interfacial transition zone is much smaller than the length of the element, so that the Young's modulus of the elements crossing the interfacial transition zone is determined as the harmonic mean of those of matrix and particle.
The strength of the element is determined by the strength of the interfacial transition zone. 
Finally, for elements with both nodes outside particles, the matrix material properties are used.
For all elements, the same structural and transport constitutive models were used with input parameters shown in Table~\ref{tab:input} according to \cite{GraWonBue10}.
\begin{savenotes}
\begin{table}
\begin{center}\caption{Input values.}
\label{tab:input}
\begin{tabular}{cccccccccccc}
\hline \\
Phase & $E$~[GPa] & $f_{\rm t}$~[MPa] & $f_{\rm c}$~[MPa] & $G_{\rm F}$~[J/m$^2$] & $A_{\rm h}$ & $\alpha$ & $\beta$ & $\psi$ & $\kappa_0$~[m$^2$] & $\xi$ \\
Matrix & 40 & 6.5 & 65 & 100 & 0.001 & 0.5 & 0.5 & 0.25 & $1 \times 10^{-19}$ & 0.001\\
ITZ & 57.1\footnote{The value E for ITZ is determined as harmonic mean of matrix and particle values.} & 3.25 & 32.5 & 50 & 0.001 & 0.5 & 0.5 & 0.25 & $1 \times 10^{-19}$ & 0.001\\
Particle & 100 & - & - & - & - & - & - & - & $1 \times 10 ^{-22}$ & -\\
\hline\\
\end{tabular}
\end{center}
\end{table}
\end{savenotes}

Three groups of analyses were carried out.
Firstly, the structural response of the matrix material subjected to mechanical loading was analysed by means of direct tension and a hydrostatic compression test using the structural periodic cell.
With these analyses key factors of the performance of the structural constitutive model were demonstrated.
This study was required since the model used here differs from the one in the previous two-dimensional study in \cite{GraWonBue10} and it is important that the numerical method provides network-independent results in tension and compression.
In the second group, periodic cells with a centrally located single particle were analysed.  
Particle restrained shrinkage was modelled by subjecting matrix and interfacial transition zones to uniform incrementally increasing eigenstrain, while keeping the force resultants of the entire specimen at zero.
The particle, which was not subjected to eigenstrain, restrained the matrix and interfacial transition zones.
Therefore, this process is called particle restrained shrinkage.
After every increment of applied eigenstrain, the permeability of the specimens was evaluated by a stationary transport analysis with a fluid pressure gradient across the specimen applied in one direction.
In these single particle analyses, the particle diameter was varied at constant particle volume fraction to investigate the influence of particle diameter on changes of permeability due to microcracking induced by restrained shrinkage.
The last group consists of coupled analyses of specimens with multiple randomly placed particles of constant diameter and volume fraction for varying specimen thickness in the direction in which the pressure gradient is applied.
With these analyses, the influence of specimen thickness on changes of permeability due to microcracking induced by particle restrained shrinkage were investigated.
In the following sections, the three groups of analyses are discussed in detail.

\subsection{Uniaxial tension and hydrostatic compression} \label{sec:mechanical}
Before investigating microcracking induced by particle restrained shrinkage, the performance of the structural network approach was investigated by means of direct tension and hydrostatic compression analyses.
For this, a cubic cell with edge length of $a=b=c=5$~cm was discretised with the periodic structural network approach using three network sizes of $d_{\rm {min}} = 8$, $4$ and $2$~mm. The number of iterations for the network generation was $N_{\rm iter} = 10000$.
The material parameters for all network elements in this cube were the one of the matrix phase shown in Table~\ref{tab:input}.

\begin{figure}
  \begin{center}
  \begin{tabular}{cc}
    \includegraphics[width=10cm]{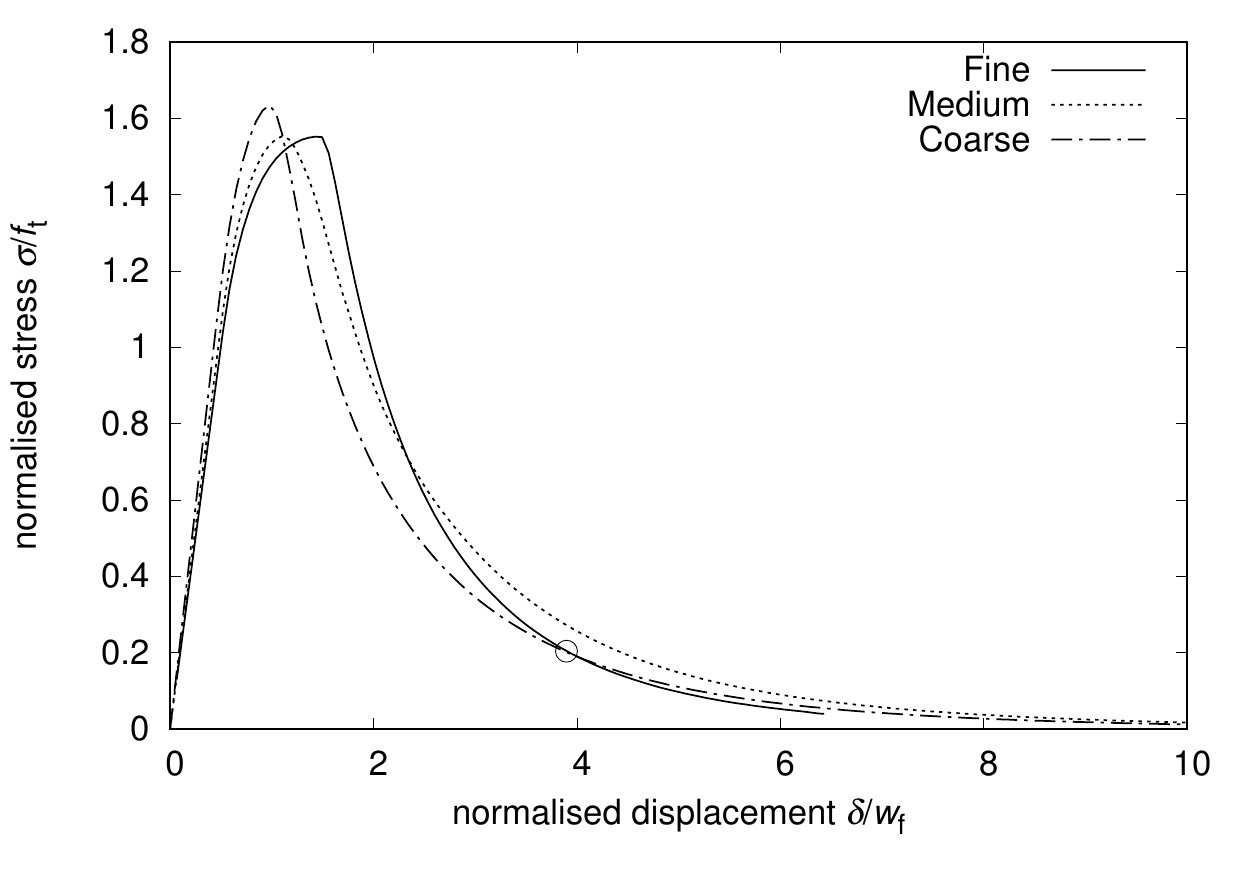} &  \includegraphics[width=6cm]{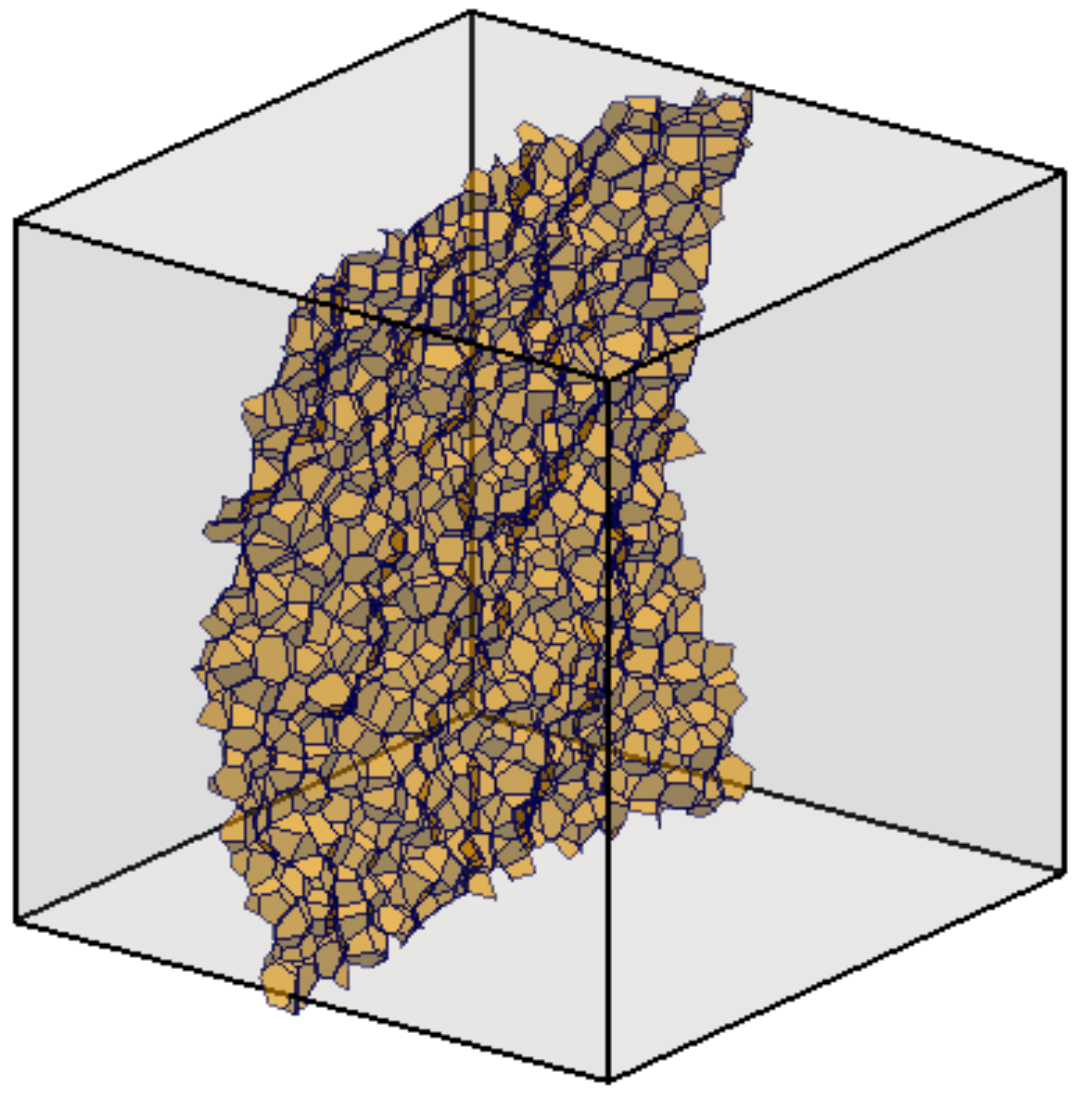}\\
    (a) & (b)
  \end{tabular}
\end{center}\vspace*{-12pt}
  \caption{Direct tension for three network sizes: a) Normalised stress versus normalised displacement in the $x$-direction, b) Crack pattern at the stage marked with a circle in a). The yellow polygons in b) show mid-cross-sections of network elements in which $\tilde{w}_{\rm c} > 10$~$\mu$m. Colours refer to the online version.}
\label{fig:tension}
\end{figure}

For direct tension, the periodic cell was subjected to monotonically increasing average axial strain $E_{\rm x}$ introduced in section~\ref{sec:structuralElement}. The average stress components corresponding to the other average strain components were set to be zero.
The resulting normalised stress-displacement curve and crack pattern for the fine network at the stage marked in the stress-displacement curve are shown in Figure~\ref{fig:tension}a~and~b, respectively. 
The structural network approach exhibits the typical response of cohesive-frictional quasibrittle materials subjected to tension in the form of softening, i.e. decreasing stress with increasing displacement (Figure~\ref{fig:tension}a), and localised deformation (Figure~\ref{fig:tension}b).
The peak stress is larger than the tensile strength input $f_{\rm t}$, because for an irregular network arrangement the individual elements are subjected to a combination of normal and shear stresses. For the present input, the shear strength is greater than the tensile strength. For shear at zero normal stress at the onset of hardening (called here $f_{\rm q}$), the elastic limit is $f_{\rm q} = 2f_{\rm t}$ for the values of the parameters $\alpha$ and $\beta$ in Table~\ref{tab:input}.

\begin{figure}
  \begin{center}
  \begin{tabular}{cc}
    \includegraphics[width=10cm]{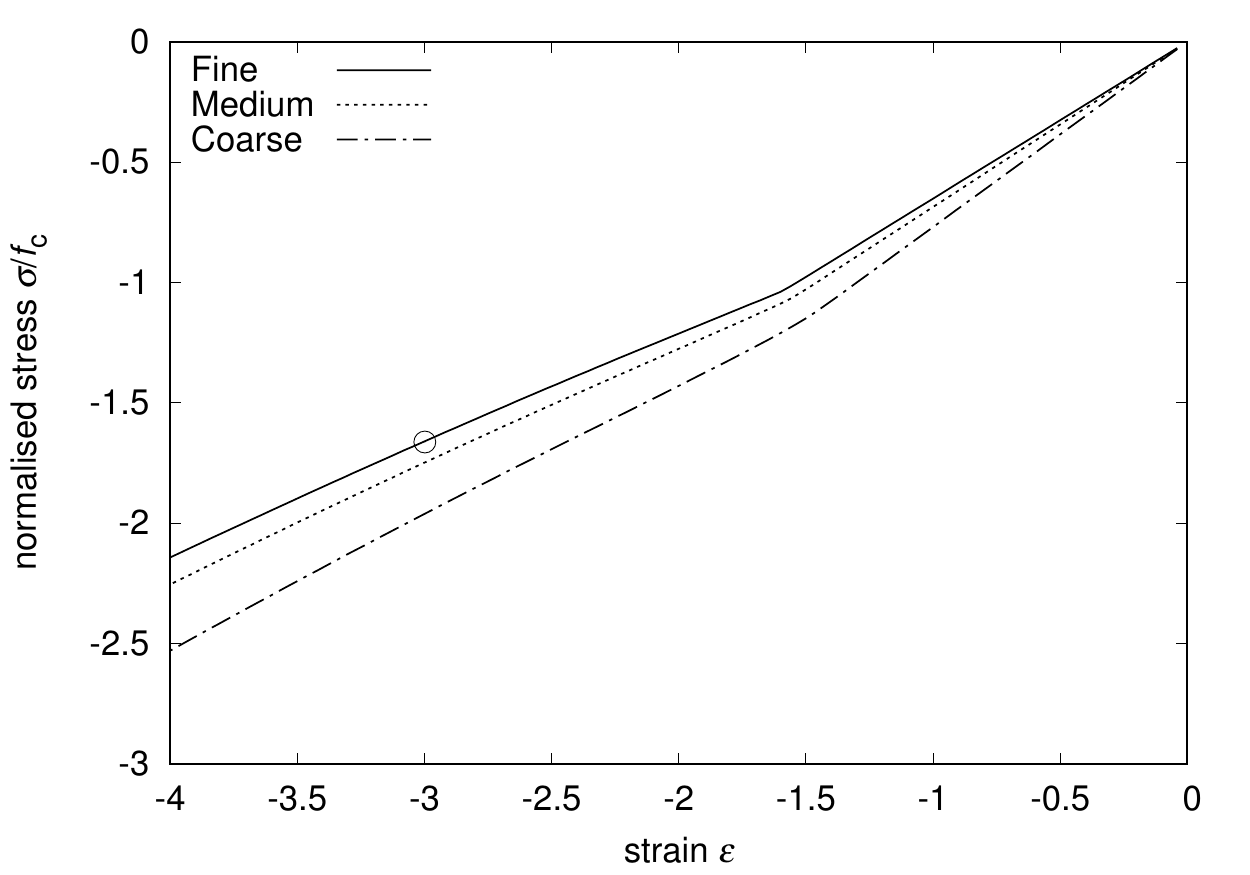} &  \includegraphics[width=6cm]{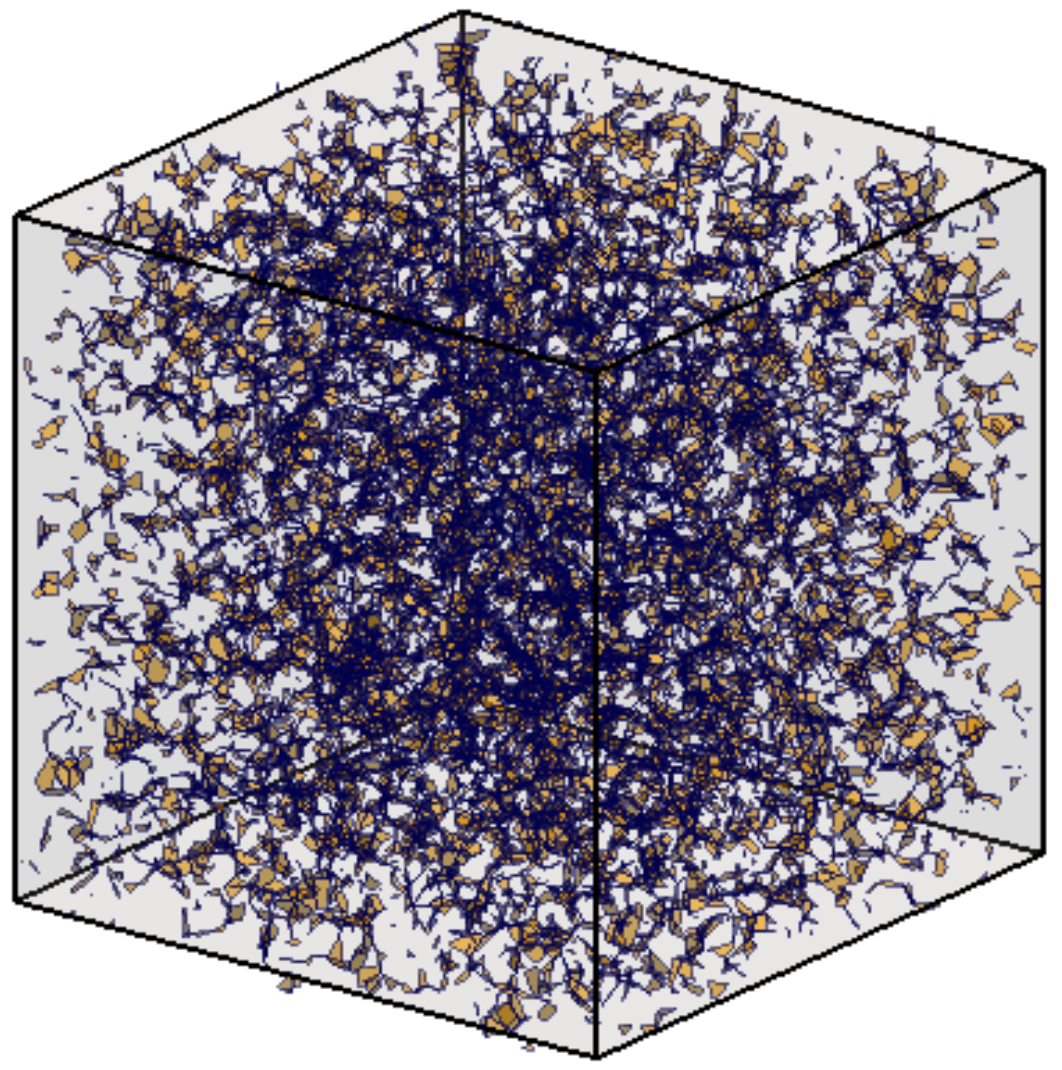}\\
    (a) & (b)
  \end{tabular}
\end{center}\vspace*{-12pt}
  \caption{Hydrostatic compression for three network sizes: a) Normalised average stress ($\sigma_{\rm v}/f_{\rm c}$) versus average strain $\varepsilon_{\rm v}$, b) Crack pattern at the stage marked with a circle in a). The yellow polygons in b) show mid-cross-sections of network elements in which $\tilde{w}_{\rm c} > 1.6$~$\mu$m. Colours refer to the online version.}
\label{fig:hydro}
\end{figure}

For hydrostatic compression, the normalised average stress $\sigma_{\rm v}/f_{\rm c}$ versus strain response and the crack patterns at the stage marked in the stress-strain curve are shown in Figure~\ref{fig:hydro}a~and~b, respectively. Here, $\sigma_{\rm v} = \left(\sigma_{\rm x} + \sigma_{\rm y} + \sigma_{\rm z}\right)/3$ and $\varepsilon_{\rm v} = \left(\varepsilon_{\rm x} + \varepsilon_{\rm y} + \varepsilon_{\rm z}\right)/3$. 
The overall response in hydrostatic compression is initially elastic followed by elasto-plastic hardening.
The deviation from the elastic response occurs at $\sigma_{\rm v} = -f_{\rm c}$, which corresponds to the onset of the yielding in the constitutive model for negative normal stress only ($\sigma_{\rm n}<0$ and $\sigma_{\rm q} = 0$).
The cracks patterns, using the definition of the crack opening in (\ref{eq:crackOpening}), are distributed within the periodic cell without showing any patterns of localised deformations, which is typical for quasi-brittle materials subjected to hydrostatic compression.
The crack opening consists only of plastic strains, since no damage occurs in these hydrostatic compression analyses. 

The two examples of direct tension and hydrostatic compression demonstrate that the structural constitutive model is capable of describing the response in tension and compression realistically.
This is important for the particle restrained shrinkage analyses in sections~\ref{sec:size}~and~\ref{sec:thickness}, in which complex tensile and compressive stress states play an important role.
The responses for both tension and compression are insensitive to the element size.
For direct tension, the irregularity of the network affects the results, because the inelastic displacements are localised in an element size dependent region.
However, global result in the form of the stress-displacement curve is insensitive to the element size.
For hydrostatic compression, the results converge with network refinement.
For large element sizes, the stiffness is overestimated because elements crossing the boundary contribute stronger to the stiffness of the cell.
Since in the analyses in sections~\ref{sec:size} the network size is varied, this insensitivity to the network size is important. 

\subsection{Size of particles} \label{sec:size}
In the second part of the study, the influence of size of inclusions on increase of permeability due to cracking was analysed.
For these analyses, a coupled periodic cubic cell ($a=b=c$ in Figure~\ref{fig:3dBox}) with a single centrally arranged particle was used.
The volume fraction was kept constant at $\rho_{\rm p} = 0.137$, while varying the particle size as $d = 4$, $8$, $16$~mm.
For the volume of a spherical particle $V_{\rm p} = \pi d^3/6$ and the cell volume $V_{\rm cell} = a^3$, the volume fraction for $n_{\rm p}$ particles is $\rho_{\rm p} = \left(n_{\rm p} V_{\rm p}\right)/V_{\rm cell}$.
Combining these expressions and solving for the specimen length gives 
\begin{equation}
a = \left(n_{\rm p} \dfrac{\pi}{6\rho_{\rm p}}\right)^{1/3} d
\end{equation}
Therefore, for constant volume fraction $\rho_{\rm p}$, the size of the periodic cell decreases with decreasing particle size.
For instance, for $n_{\rm p} = 1$ and $d=16$~mm, the specimen length results in $a = 25$~mm.
For the discretisation of the network, the ratio of the size of the cell and minimum distance was chosen as $a/d_{\rm{min}} = 12.5$ for all particle sizes investigated.
Thus, the average network element length decreases with decreasing particle size.
This change of element size should not affect the results strongly, as it was shown in section~\ref{sec:mechanical}.
For all analyses $N_{\rm iter} = 10000$ was used.

For the coupled analyses, a uniform shrinkage strain of $\varepsilon_{\rm s}=-0.5$\% was applied in 100 increments to matrix and interfacial transition zone.
The particles were not subjected to the shrinkage strain. 
For each particle size, ten analyses with random network generations were performed. 
After every increment of shrinkage strain, the permeability was determined by applying a unit fluid pressure gradient in the $y$-direction  ($\Delta P_{\rm y}/L_{\rm y} = 1$).
Here, $\Delta P_{\rm y}$ and $L_{\rm y} = a = b= c$ are the fluid pressure difference and the length in the $y$-direction, respectively.
The total flow $Q_{\rm y}$ in the $y$-direction resulting from this fluid pressure gradient was used to determine the macroscopic permeability component of the cell in the $y$-direction as
\begin{equation}\label{eq:permeabilityComponent}
\kappa_{\rm yy} = \dfrac{Q_{\rm y} \mu}{A_{\rm y} \rho \Delta P_{\rm y}/L_{\rm y}} 
\end{equation}
where $A_{\rm y} = a\times b$ is the cross-sectional area in $y$-direction (Figure~\ref{fig:3dBox}).
The permeability component in (\ref{eq:permeabilityComponent}), which is one of nine components of the matrix of permeability \cite{NilLarLun14}, was used here to assess the influence of particle size on permeability for microcracking induced by particle restrained shrinkage.

\begin{figure}
  \begin{center}
  \begin{tabular}{c}
    \includegraphics[width=12cm]{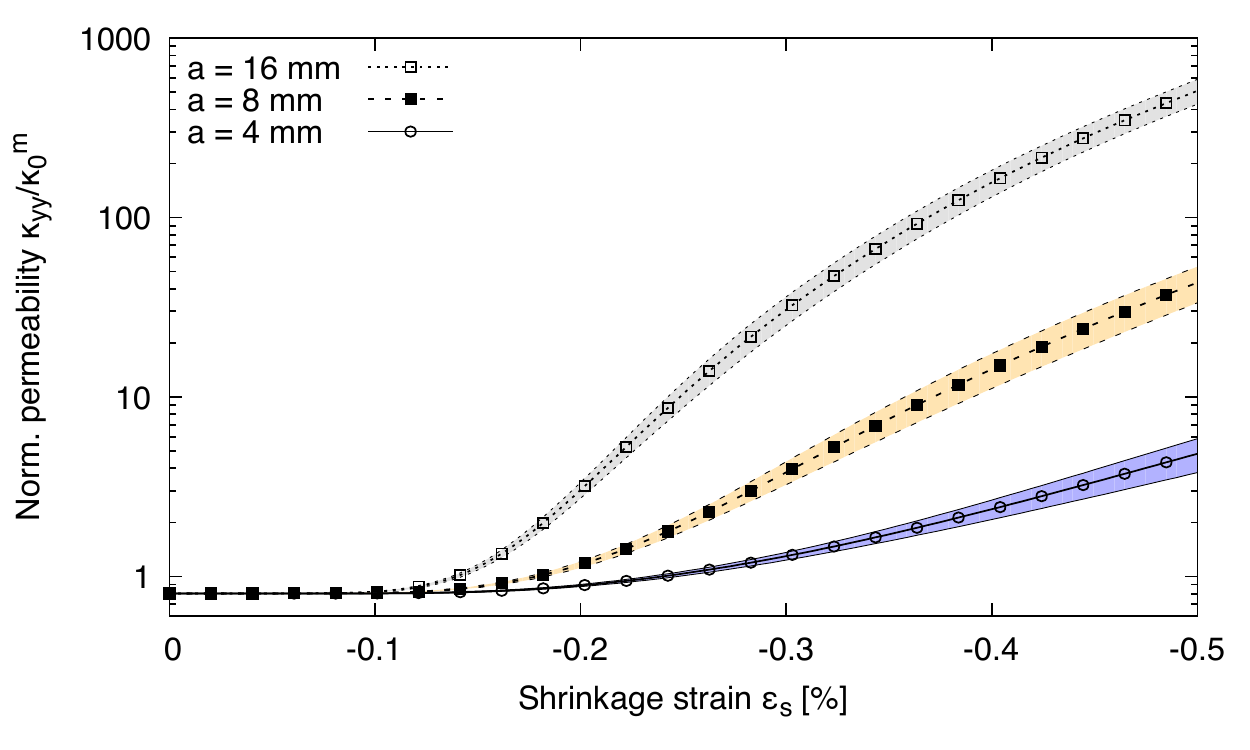}
  \end{tabular}
\end{center}\vspace*{-12pt}
  \caption{Particle size: Permeability $\kappa_{\rm yy}$ normalised by the intrinsic permeability of the matrix $\kappa_{0}^{m}$ versus shrinkage strain $\varepsilon_{\rm s}$ for three particle diameters at constant particle volume fraction.}
\label{fig:size}
\end{figure}

\begin{figure}
  \begin{center}
  \begin{tabular}{ccc}
    \includegraphics[width=5cm]{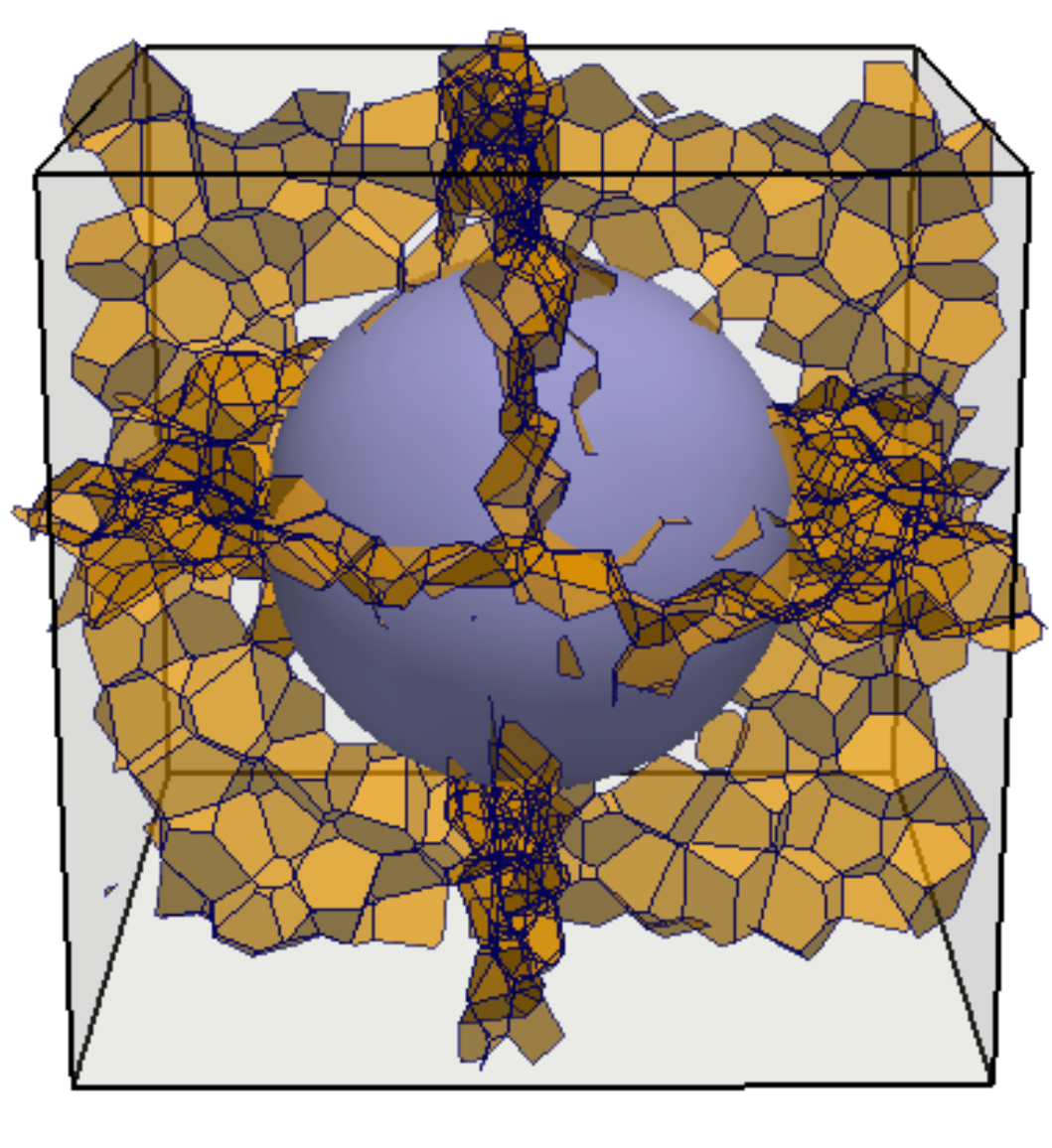} & \includegraphics[width=5cm]{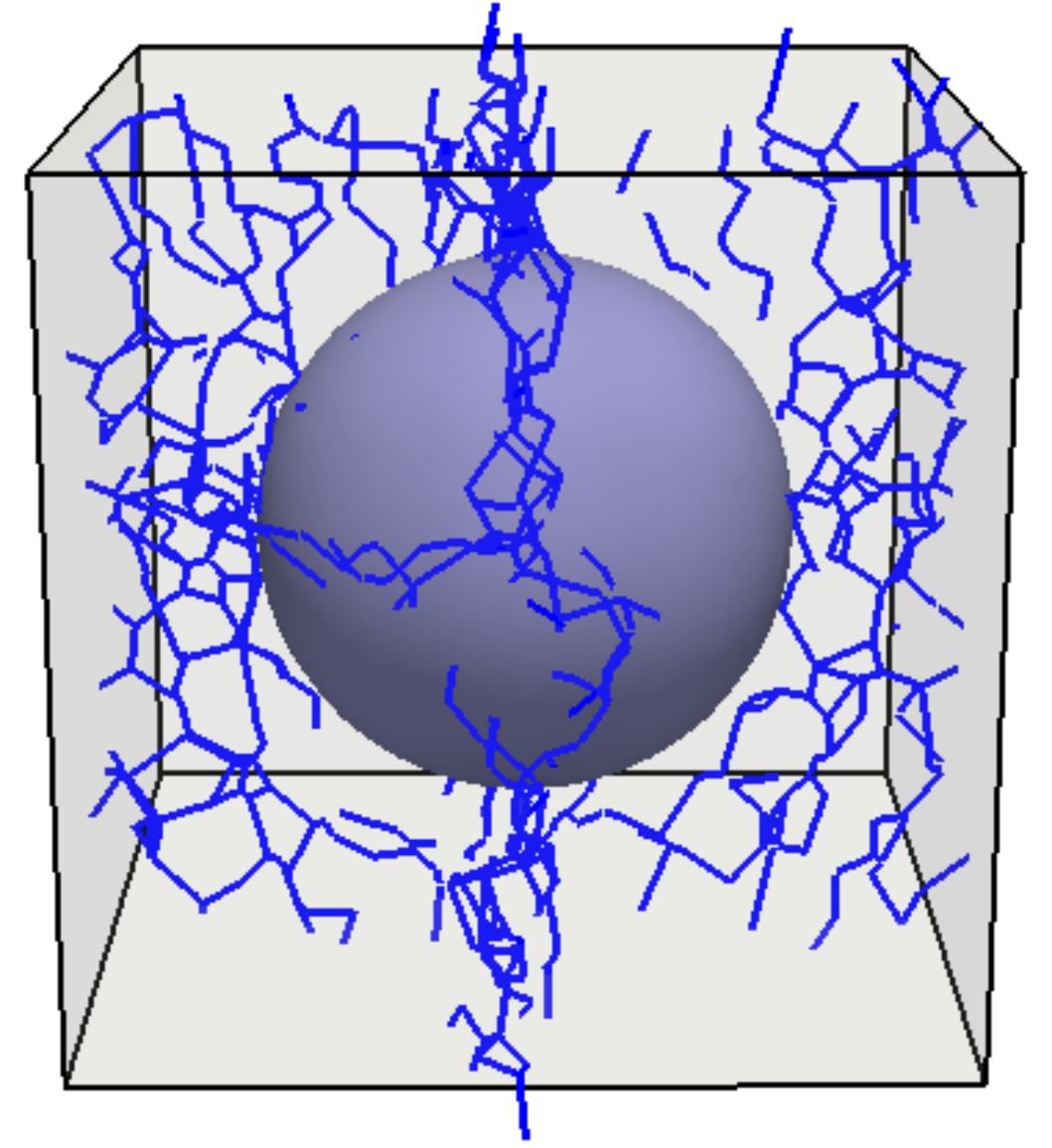}\\
    (a) & (b)
  \end{tabular}
\end{center}\vspace*{-12pt}
  \caption{Particle size: (a) Crack pattern and (b) flow paths for $d=16$~mm and $\rho = 0.137$ at the final increment of shrinkage strain. Yellow polygons in a) show mid-cross-sections of elements in which crack openings increase at this stage of analysis and $\tilde{w}_{\rm c} > 10$~$\mu$m. Blue lines in b) show transport elements in which the flow is greater than the threshold $Q_0$. Colours refer to the online version.}
\label{fig:singleCrack}
\end{figure}

In Figure~\ref{fig:size}, the mean of the permeability $\kappa_{\rm yy}$ normalised by the intrinsic permeability of the matrix $\kappa_0^{m}$ versus shrinkage strain $\varepsilon_{\rm s}$ of ten random analyses is shown. 
The areas next to the curves show plus/minus one standard deviation.
The scatter originates from the irregularity of the structural and transport networks.
At early stages of the analyses ($\varepsilon_{\rm s}>-0.1$~\%), microcracking due to particle restraint does not occur, so that there is no visible influence of particles size on permeability on the log-scale used in Figure~\ref{fig:size}.
Once cracking has been initiated ($\varepsilon_{\rm s}<-0.1$~\%), permeability is strongly influenced by the particle size at constant particle volume fraction.
The greater the particle is, the greater is the increase of permeability. 
This strong dependence of permeability on particle size at constant volume fraction is explained by the crack patterns which are generated by the shrinkage of matrix and interfacial transition zone and the restraint that the particle provides.
Crack patterns for the largest particle size ($d=16$~mm) are shown in Figure~\ref{fig:singleCrack}a at the final increment of shrinkage strain.
In this figure, cracks are visualised by yellow polygons representing mid-cross-sections of elements in which the equivalent crack width $\tilde{w}_{\rm c}$ defined in (\ref{eq:equivCrack}) is greater than 10~$\mu$m.
Colours refer to the online version.
For all analyses of periodic cells with a single particle size, the shrinkage strain applied to matrix and interfacial transition zone results in overall regular localised crack patterns with three distinct crack planes aligned with the directions of the Cartesian coordinate system used for the periodicity of the cell.
In an earlier two-dimensional study in \cite{GraWonBue10}, these type of regular crack patterns were obtained by placing multiple layers of inclusions in a regular pattern in a bigger specimen.
Here, because a cell with periodic boundary conditions is used, these characteristic crack patterns for a regular inclusion arrangement are obtained for a cell with a single inclusion.

The crack openings depend strongly on particle size. The greater the particle size is, the greater is the volume of material associated with this particle, which will be subjected to the eigenstrain and which will crack if the eigenstrain is sufficiently large.
Therefore, the greater the particle size, the greater is the crack opening and the smaller is the crack length.
This reasoning is in agreement with the two-dimensional numerical results reported in \cite{GraWonBue10}.
This particle size dependent crack opening results in a strong dependence of conductivity on particle size, since the cubic law in (\ref{eq:crackDiff}) was used to related crack opening to permeability due to cracking. Therefore, the increase in crack openings dominates the decrease in crack length.

In addition to the crack patterns, the main flow paths through the crack network is visualised in Figure~\ref{fig:singleCrack}b by showing transport elements in blue in which the flow is greater than the threshold $Q_0 = 2.5\times 10^{-16}$~kg/s.
This value is equal to the entire flow through a cell of the same cross-section made of undamaged matrix material and subjected to a unit pressure gradient.
In Figure~\ref{fig:singleCrack}b, it can be seen that the flow paths are orientated preferentially in the vertical direction.
Two of the crack planes in Figure~\ref{fig:singleCrack}a are orientated so that they result in an increase of the mass transport in the $y$-direction.
Most of the transport elements exceeding the threshold are located on these two crack planes.
The strong dependence of permeability on particle size in Figure~\ref{fig:size} is in agreement with the two-dimensional results in \cite{GraWonBue10}.

\subsection{Specimen thickness} \label{sec:thickness}
\begin{figure}
  \begin{center}
  \begin{tabular}{c}
    \includegraphics[width=12cm]{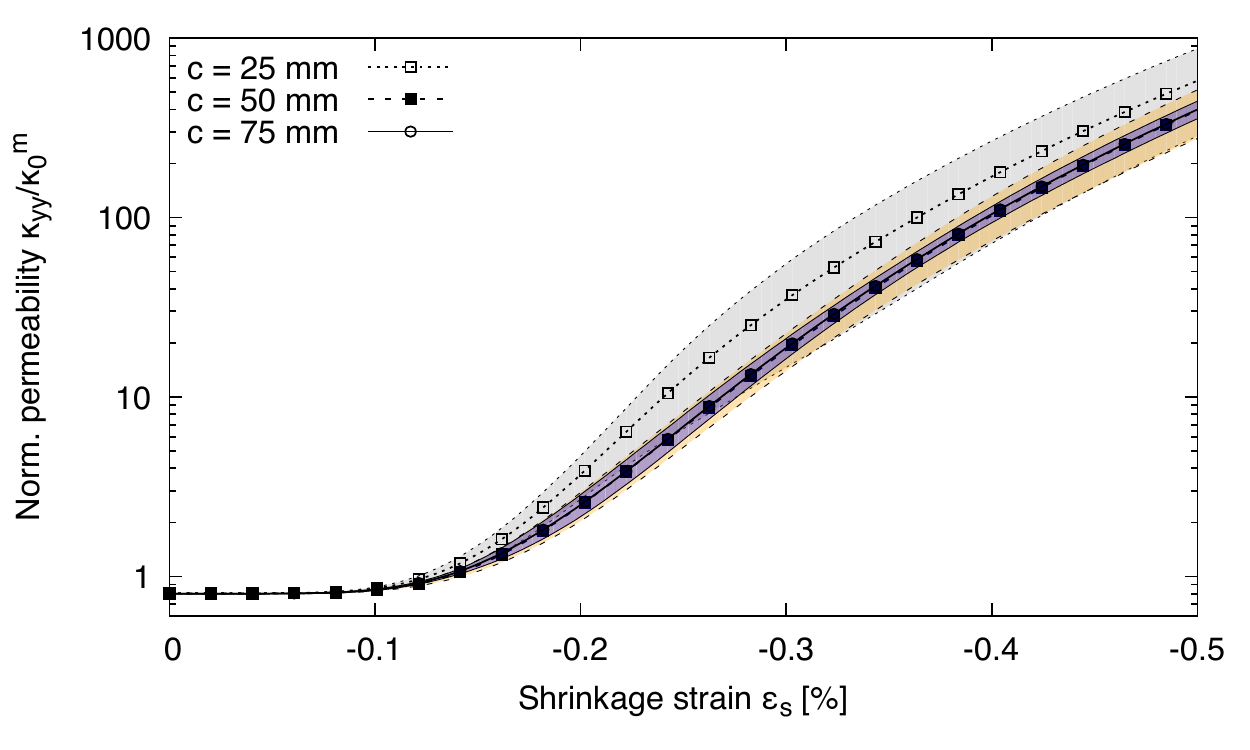}
  \end{tabular}
\end{center}\vspace*{-12pt}
  \caption{Specimen thickness: Permeability $\kappa_{\rm yy}$ normalised by the intrinsic permeability of the matrix $\kappa_{\rm 0}^{m}$ versus shrinkage strain $\varepsilon_{\rm s}$ for three specimen thicknesses at constant particle volume fraction and particle diameter.}
\label{fig:thickPermeability}
\end{figure}

\begin{figure}
  \begin{center}
  \begin{tabular}{ccc}
    \includegraphics[width=5cm]{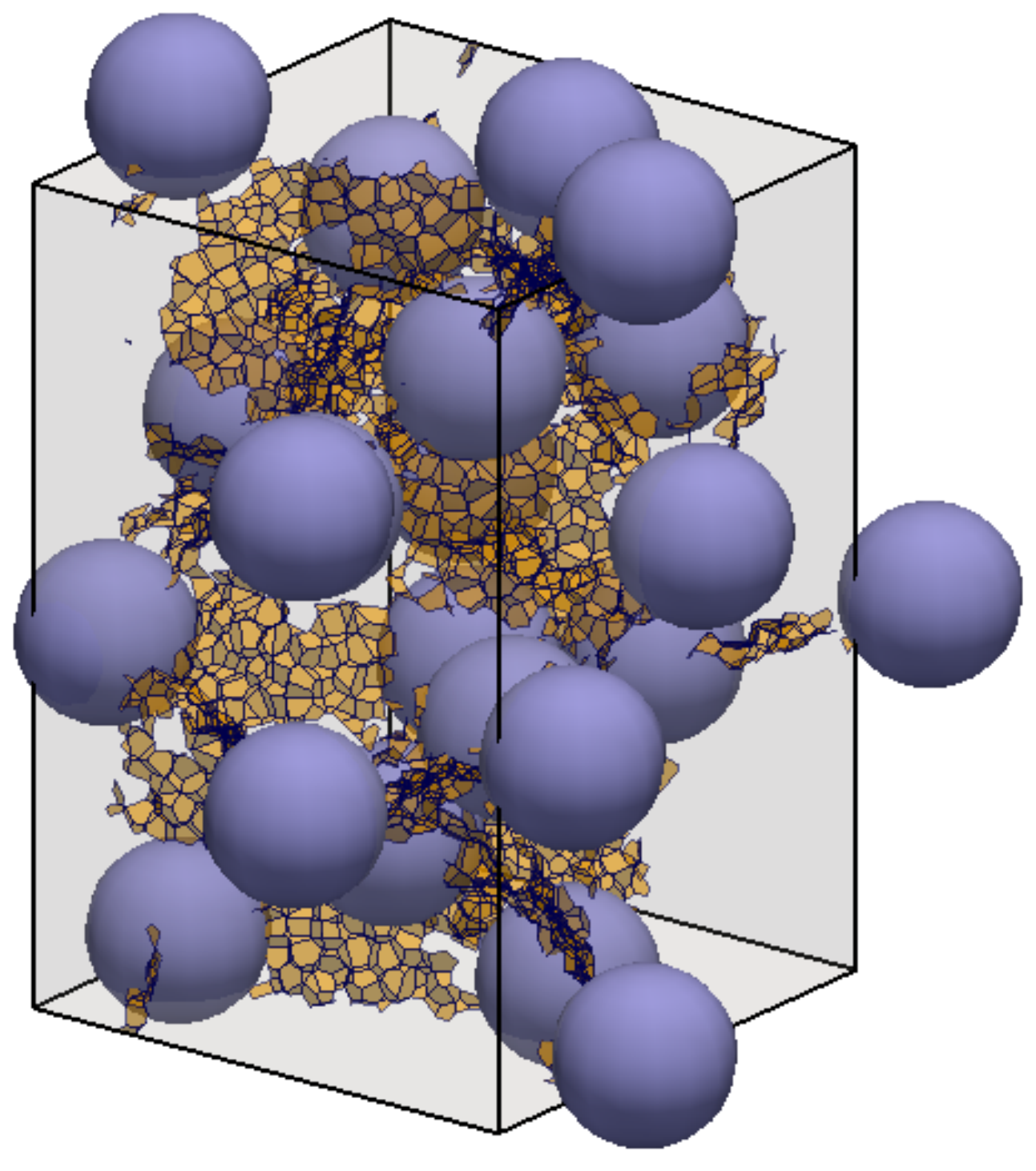} & \includegraphics[width=5cm]{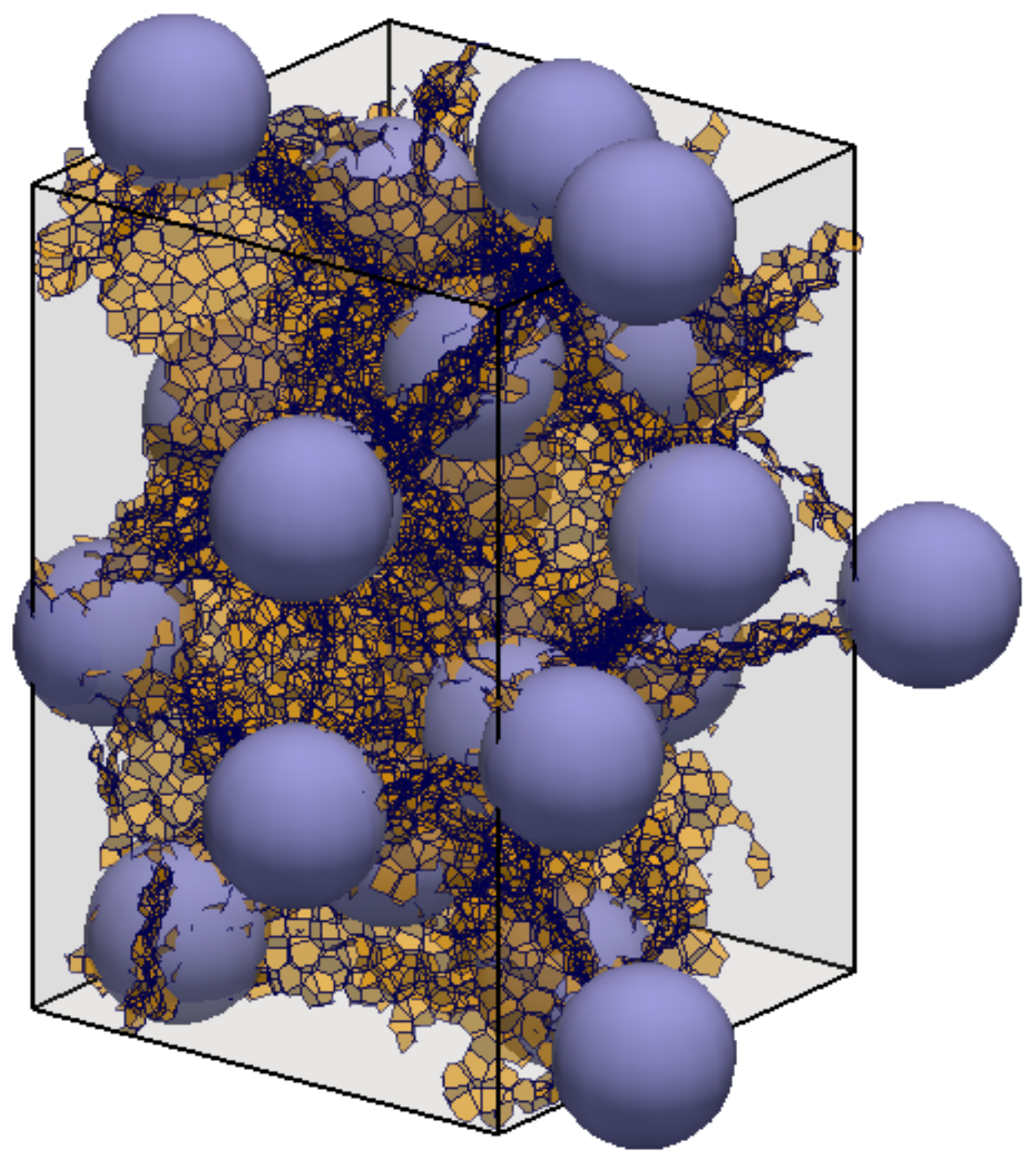} & \includegraphics[width=5cm]{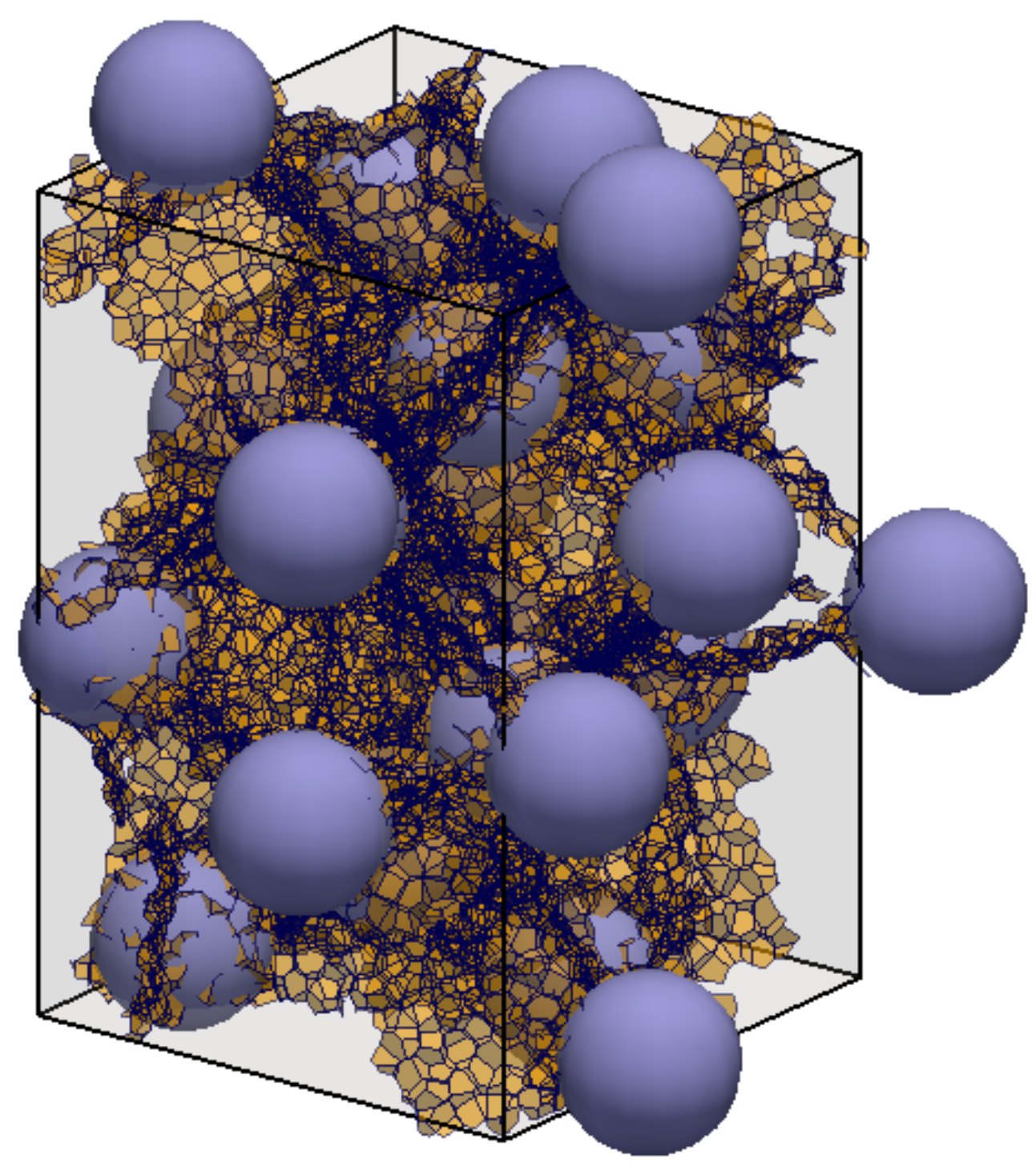}\\
    (a) & (b) & (c)
  \end{tabular}
\end{center}\vspace*{-12pt}
  \caption{Specimen thickness: Crack patterns for the thick specimen ($c=75$~mm) for three stages of shrinkage strain of (a) $\varepsilon_{\rm s} = -0.25$, (b) $-0.375$ and (c) $-0.5$~\%. Particles are shown by blue spheres and the cracks are shown as yellow polygons representing mid-cross-sections of structural elements in which crack openings increase at this stage of the analysis and are greater than 10~$\mu$m. Colours refer to the online version.}
\label{fig:crackEvolv}
\end{figure}

In the third part, the influence of specimen thickness on random particle arrangements in the direction of the pressure gradient was investigated for microcracking induced by particle restrained shrinkage.
The aim of this part of the study was to investigate if potentially disconnected crack networks result in reduction of permeability with increasing specimen thickness.
The cross-section of the rectangular periodic cell was chosen as $a=b=50$~cm and the specimen thickness was varied as $c=25$, $50$ and $75$~mm (Figure~\ref{fig:3dBox}).
Furthermore, particle diameter and density were chosen as $d=16$~mm and $\rho_{\rm p} = 0.137$, respectively.
The minimum distance for the background network was chosen as $d_{\rm{min}} = 2$~mm.
For each specimen thickness, ten analyses with random periodic particle and network arrangements were carried out using the coupled network approach presented in section~\ref{sec:method}.
As in the single particle analyses in section~\ref{sec:size}, a shrinkage strain of $\varepsilon_{\rm s} = -0.5$~\% was applied incrementally to matrix and interfacial transition zone elements.
After every increment, the permeability component $\kappa_{\rm yy}$ was determined as explained in (\ref{eq:permeabilityComponent}).
The mean of the permeability of ten random analyses versus the shrinkage strain is presented in Figure~\ref{fig:thickPermeability} in the form of lines with symbols for the three specimen thicknesses. 
Note that the mean results for $c=50$~and~$75$~mm are almost indistinguishable.
The coloured areas around the mean curves represent plus/minus one standard deviation.
The permeability increases strongly with increasing shrinkage strain, as it was already observed for the single particle analyses.
The strong increase of permeability is the result of the creation of random crack networks.
These are more complex than the regular patterns obtained in section~\ref{sec:size}.
However, the permeability at the final stage is similar for the random and regular particle arrangements.
In Figure~\ref{fig:crackEvolv}, the crack patterns of one of the thick specimens ($c=75$~mm) is shown for three stages of applied shrinkage strain ($\varepsilon_{\rm s} = -0.25$, $-0.375$ and $-0.5$~\%).
The blue spheres indicate the position of the particles.
The yellow polygons show mid-cross-sections of elements in which the crack opening increases at this stage of analysis and is greater than 10~$\mu$m.
Colours refer to the online version.
The spacing between the crack planes is determined by the size of the particles and their spatial arrangement.
For all three stages of applied shrinkage strains, the crack network connects randomly placed particles.
Very similar observations were made in the two-dimensional study in \cite{GraWonBue10}.
The greater the shrinkage strain, the denser is the crack network and the greater are the crack openings.

\begin{figure}
  \begin{center}
  \begin{tabular}{ccc}
    \includegraphics[width=5cm]{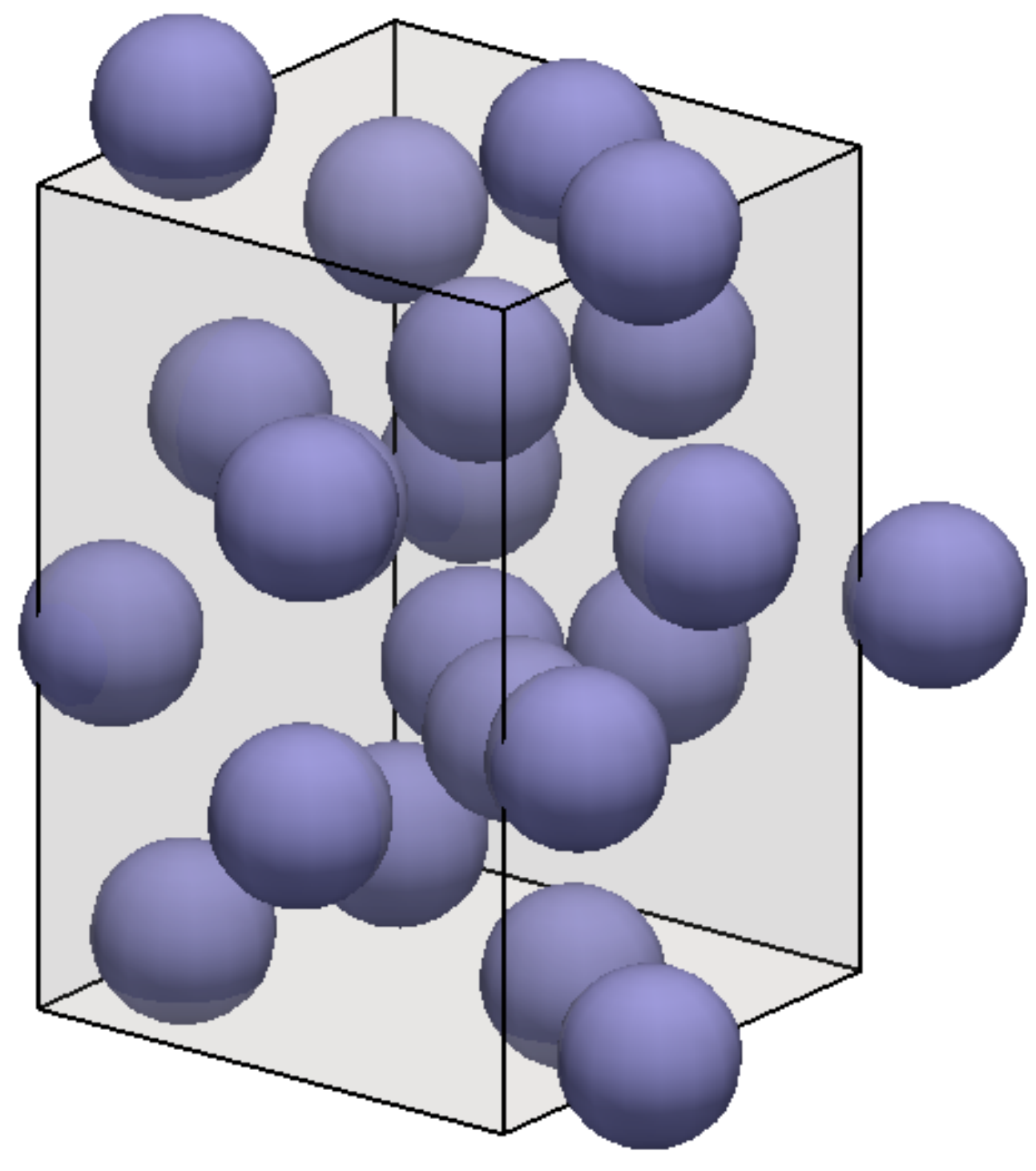} & \includegraphics[width=5cm]{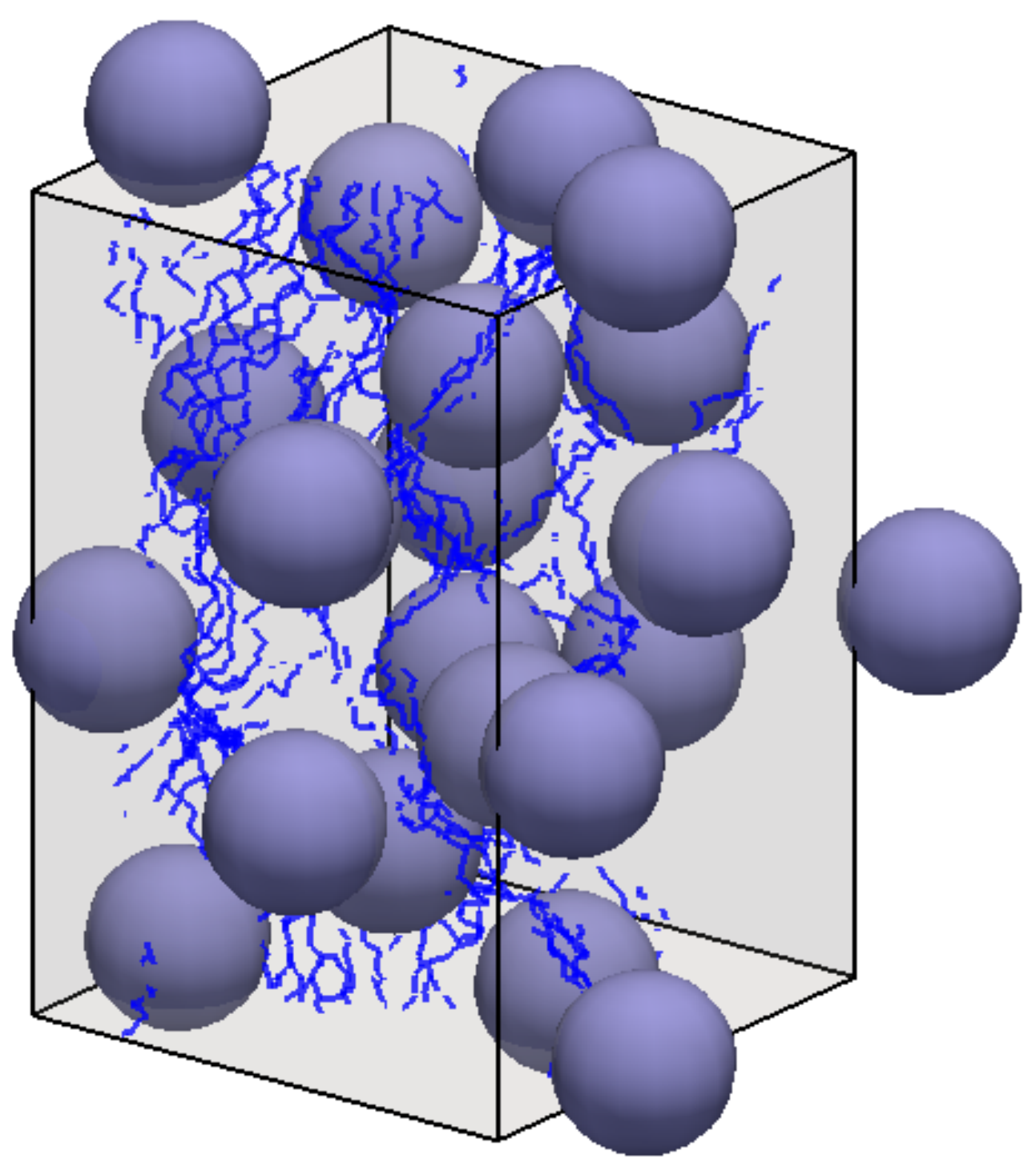} & \includegraphics[width=5cm]{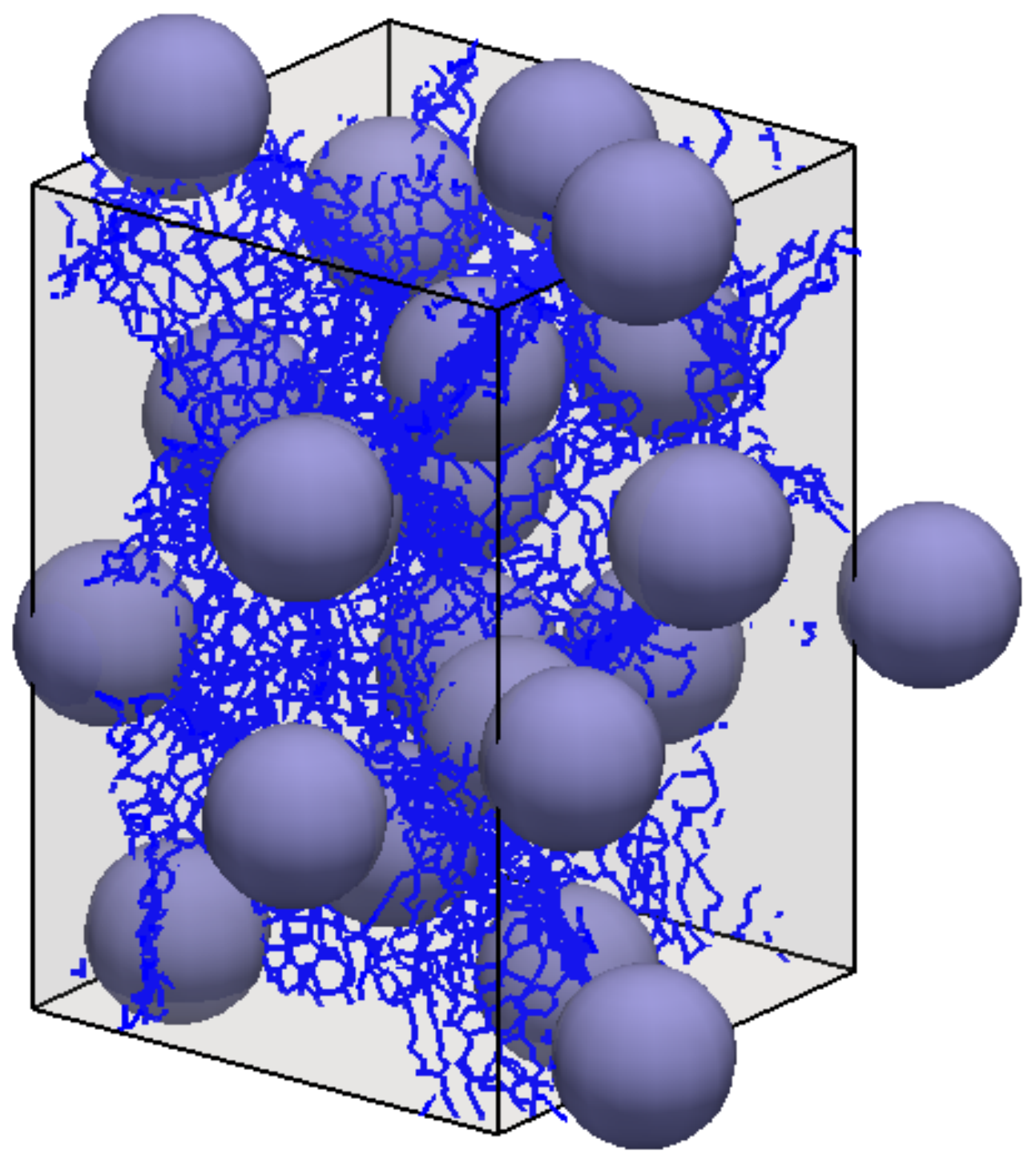}\\
(a) & (b) & (c)
  \end{tabular}
\end{center}\vspace*{-12pt}
  \caption{Specimen thickness: Flow patterns in the thick specimen ($c=75$~mm) for three stages of shrinkage strain of (a) 0.25\%, (b) 0.375\% and (c) 0.5\% applied to the matrix. Particles are shown as blue spheres. Blue lines show transport elements in which the flow is greater than $Q_0$.}
\label{fig:flowEvolv}
\end{figure}

In addition, the flow network is shown in Figure~\ref{fig:flowEvolv}.
Blue lines indicate transport elements in which the flow is greater than $Q_0$, which is the threshold used earlier for the single particle analyses.
At the first stage in Figure~\ref{fig:flowEvolv}a, none of the transport elements exhibits flow greater than $Q_0$.
The crack openings of the crack networks in Figure~\ref{fig:crackEvolv}a are, while already localised, not large enough to increase the flow through the transport element sufficiently to exceed the threshold $Q_0$.
For the second stage in Figure~\ref{fig:crackEvolv}b, a network of transport elements exceeding the threshold is visible.
These transport elements are located on the crack planes shown in Figure~\ref{fig:crackEvolv}b.
Not all transport elements on crack planes exhibit high flow, because of the variation of the crack openings of individual elements.
In the final stage, the majority of transport elements located on crack planes shown in Figure~\ref{fig:crackEvolv}b conduct flow greater than $Q_0$. 
At this stage, the crack network is fully developed and the cracks have opened up so much that transport elements along the crack planes provide the majority of the flow through the specimen.

\begin{figure}
  \begin{center}
  \begin{tabular}{ccc}
    \includegraphics[width=4cm]{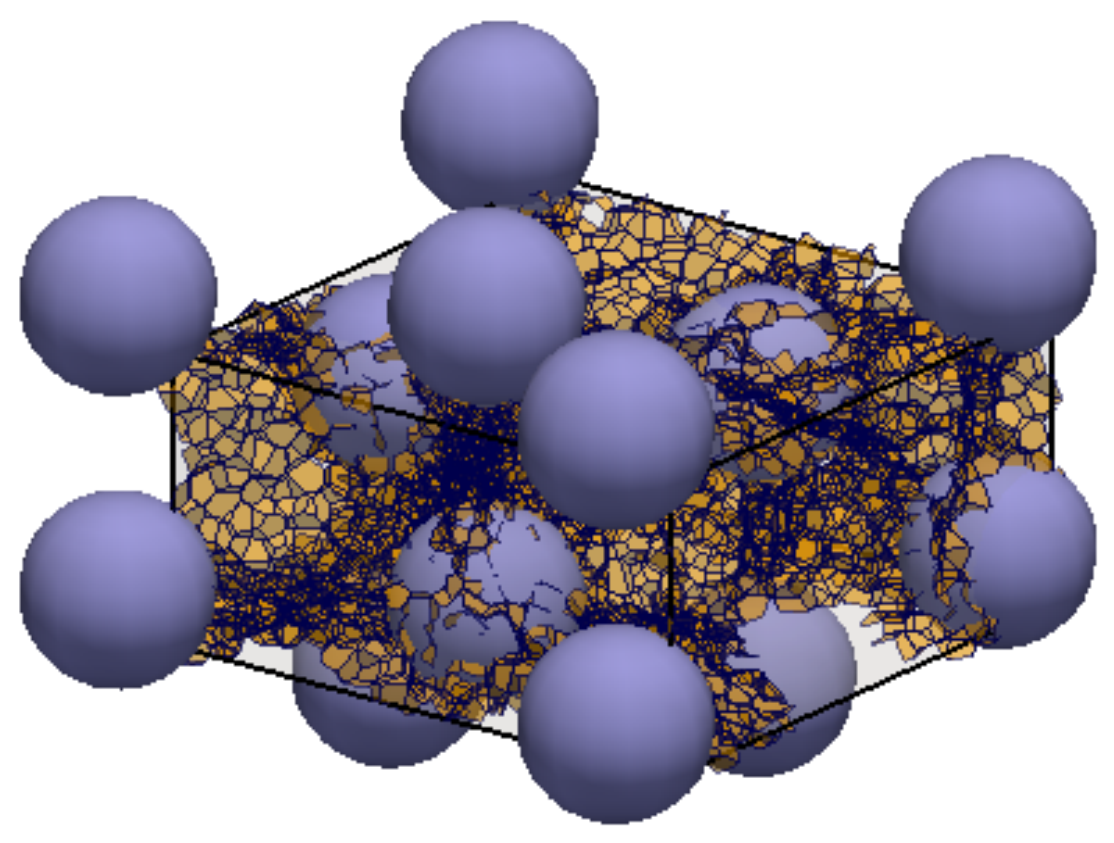} & \includegraphics[width=4cm]{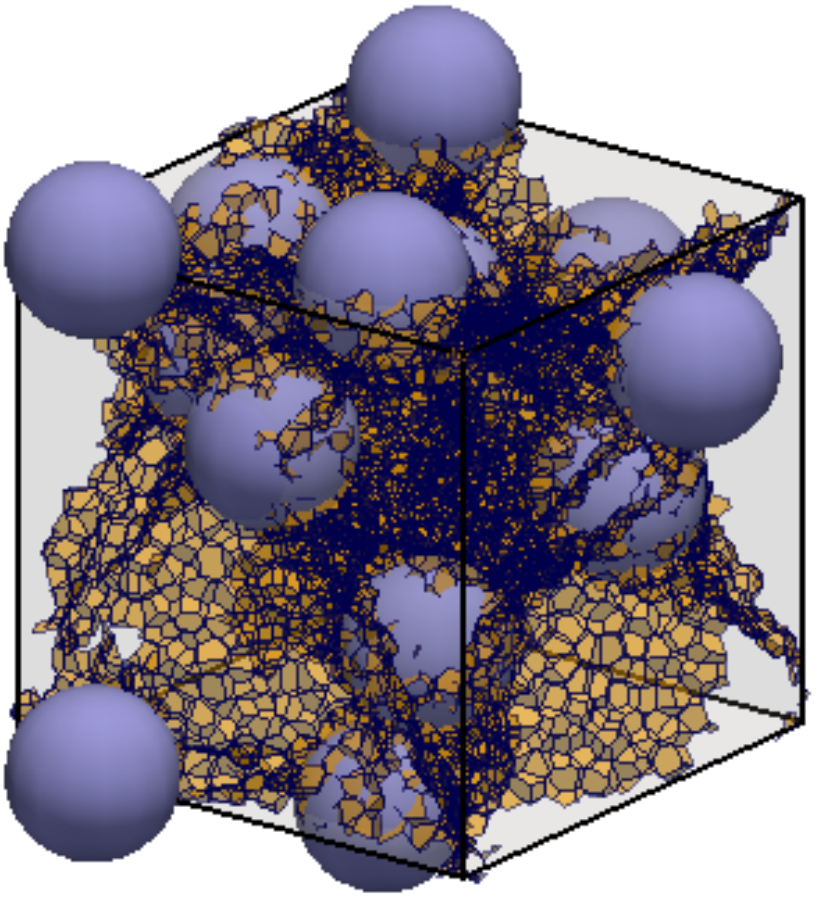} & \includegraphics[width=5cm]{./figLongCrackStep100.pdf}\\
(a) & (b) & (c)
  \end{tabular}
\end{center}\vspace*{-12pt}
  \caption{Specimen thickness: Crack patterns for specimens with three thicknesses ($c=25$, $50$, $75$~mm) for the final stage of shrinkage strain ($-0.5$~\%). Particles are shown by blue spheres and the cracks are shown as yellow polygons representing mid-cross-sections of structural elements in which crack openings increase at this stage of the analysis and are greater than 10~$\mu$m. Colours refer to the online version.}
\label{fig:crackThickness}
\end{figure}

The specimen thickness has overall only a small influence on the increase of permeability compared to the influence of the particle size and shrinkage strain.
The mean permeability for $c=50$~and~$75$~mm are almost identical.
Only for the specimen with $c=25$~mm, a greater permeability than for $c=50$ and $75$~mm was obtained.
This difference is smaller than the standard deviation of the analyses with $c=25$~mm.
The greater the thickness is, the smaller is the standard deviation of the permeability, because some of the irregularities of the crack planes are averaged out along the specimen thickness.
Crack patterns for specimens of three different thicknesses are shown in Figure~\ref{fig:crackThickness} for the final stage ($\varepsilon_{\rm s} = -0.5$).
For all three thicknesses, crack patterns connecting the particles are fully formed.
Qualitatively, there is little difference between these crack patterns.

\begin{figure}
  \begin{center}
  \begin{tabular}{ccc}
    \includegraphics[width=5cm]{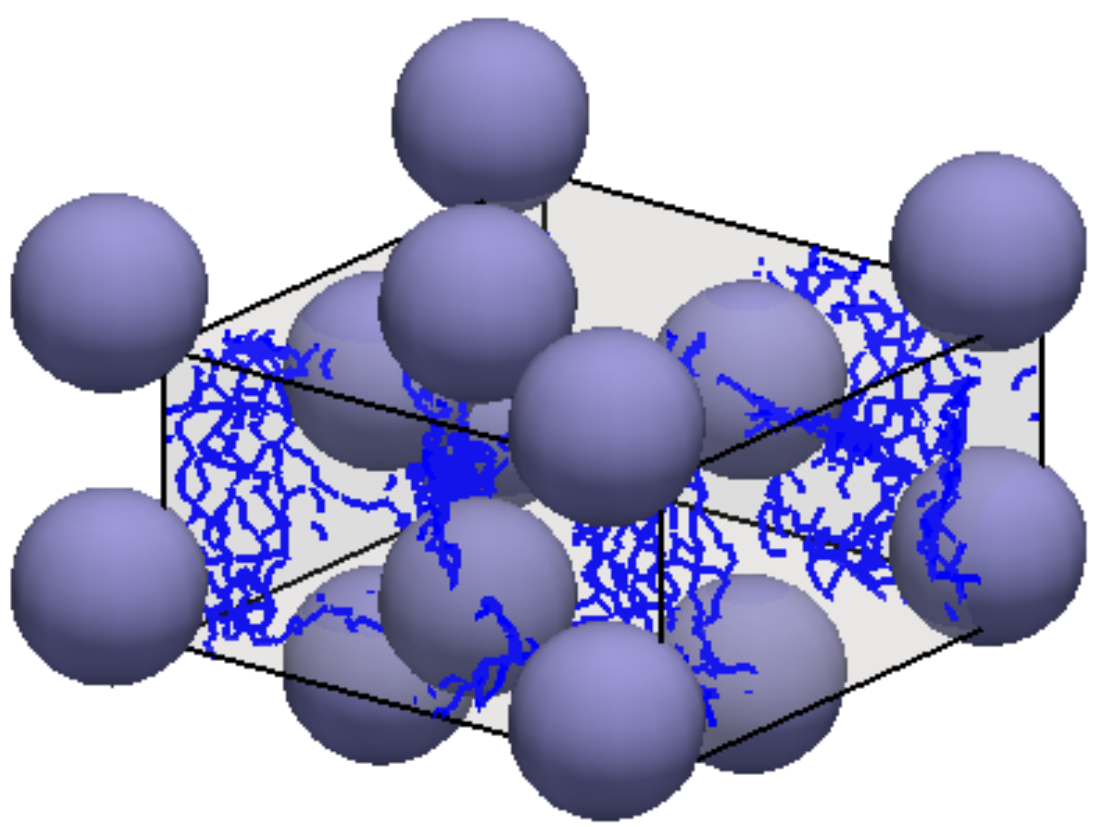} & \includegraphics[width=4cm]{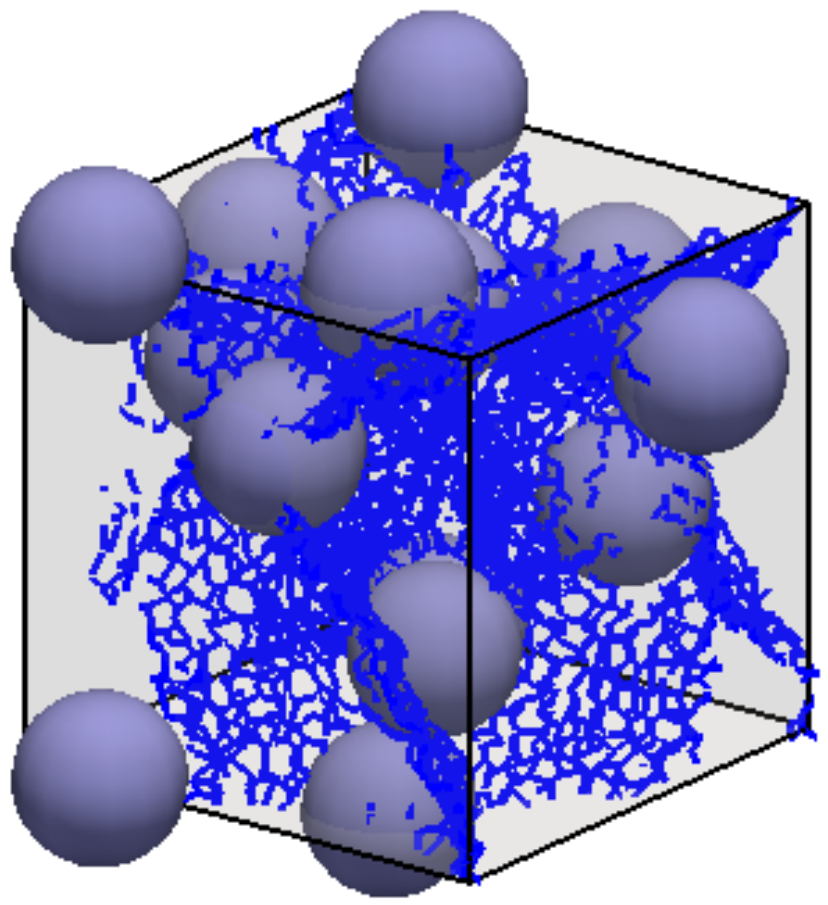} & \includegraphics[width=5cm]{figLongFlowStep100.pdf}\\
(a) & (b) & (c)
  \end{tabular}
\end{center}\vspace*{-12pt}
  \caption{Specimen thickness: Flow patterns for specimens with three thicknesses ($c=25$, $50$, $75$~mm) for the final stage of shrinkage strain ($-0.5$~\%). Particles are shown by blue spheres. Blue lines show transport elements in which the flow is greater than $Q_0$.}
\label{fig:flowThickness}
\end{figure}

In Figure~\ref{fig:flowThickness}, the network of transport elements in which the flow is greater than the threshold $Q_0$ is shown for a shrinkage strain of $\varepsilon_{\rm s} = -0.5$.
For all three specimens, the network of transport elements exceeding this threshold have a very similar structure.
The high flow elements are positioned on the crack planes as shown in Figure~\ref{fig:crackThickness}.
For the network with the smallest specimen thickness ($c=25$~mm) in Figure~\ref{fig:flowThickness}a, the number of elements with high flow is smaller than for the other two thicknesses in Figure~\ref{fig:flowThickness}b~and~c. This appears to be in contradiction to the results in Figure~\ref{fig:thickPermeability}, in which the specimen with the smallest thickness exhibits the greatest increase in conductivity due to cracking. Nevertheless, the crack and flow patterns in Figures~\ref{fig:crackThickness} and \ref{fig:flowThickness} are only one of ten random particle arrangements.
These illustrations are useful for assessing the type of crack and flow patterns, but cannot be used for a quantitative comparison analyses with different thicknesses.
For such a comparison, the mean values shown in Figure~\ref{fig:thickPermeability} should be used.
In addition, for the smallest specimen thickness, the highest standard deviation was reported. 

\section{Conclusions}
A three-dimensional coupled structural transport network model was used for the analysis of microcracking in heterogeneous materials due to particle restrained shrinkage.
A new coupled periodic network approach has been proposed, which combines periodic displacement/pressure-gradient conditions with periodic structural and transport networks.
This new approach was applied to investigate the influence of particle size and specimen thickness on permeability for particle restrained shrinkage.
The results of the three-dimensional analyses show that a change of particle size at constant volume fraction has a very strong influence on the increase of permeability due to microcracking for the case of particle restrained shrinkage.
  The influence of a change specimen thickness on increase of permeability for a pressure gradient in the direction of the changed specimen thickness was shown to be small in comparison.

This study was limited to shrinkage of matrix and ITZ applied uniformly across the periodic cells. In many applications, the shrinkage strain will be nonuniform through the specimen because of for instance a humidty gradient away from the specimen surface \citep{HavJir16}. This influence of nonuniform shrinkage strain on microcracking and conductivity changes will be investigated in future studies.

\section*{Acknowledgements}
The numerical analyses were performed with the nonlinear analyses program OOFEM \cite{Pat12} extended by the present authors.
The authors acknowledge funding received from the UK Engineering and Physical Sciences Research Council (EPSRC) under grant EP/I036427/1 and funding from Radioactive Waste Management Limited (RWM) (http://www.nda.gov.uk/rwm), a wholly-owned subsidiary of the Nuclear Decommissioning Authority. RWM is committed to the open publication of such work in peer reviewed literature, and welcomes e-feedback to rwmdfeedback@nda.gov.uk.

\bibliographystyle{elsarticle-harv}
\bibliography{general}

\begin{thebibliography}{26}
\expandafter\ifx\csname natexlab\endcsname\relax\def\natexlab#1{#1}\fi
\providecommand{\url}[1]{\texttt{#1}}
\providecommand{\href}[2]{#2}
\providecommand{\path}[1]{#1}
\providecommand{\DOIprefix}{doi:}
\providecommand{\ArXivprefix}{arXiv:}
\providecommand{\URLprefix}{URL: }
\providecommand{\Pubmedprefix}{pmid:}
\providecommand{\doi}[1]{\href{http://dx.doi.org/#1}{\path{#1}}}
\providecommand{\Pubmed}[1]{\href{pmid:#1}{\path{#1}}}
\providecommand{\bibinfo}[2]{#2}
\ifx\xfnm\relax \def\xfnm[#1]{\unskip,\space#1}\fi
\bibitem[{Akhavan et~al.(2012)Akhavan, Shafaatian and Rajabipour}]{AkhShaRaj12}
\bibinfo{author}{Akhavan, A.}, \bibinfo{author}{Shafaatian, S.M.H.},
  \bibinfo{author}{Rajabipour, F.}, \bibinfo{year}{2012}.
\newblock \bibinfo{title}{Quantifying the effects of crack width, tortuosity,
  and roughness on water permeability of cracked mortars}.
\newblock \bibinfo{journal}{Cement and Concrete Research} \bibinfo{volume}{42},
  \bibinfo{pages}{313--320}.
\bibitem[{Athanasiadis(2017)}]{Ath17}
\bibinfo{author}{Athanasiadis, I.}, \bibinfo{year}{2017}.
\newblock \bibinfo{title}{Hydro-mechanical network modelling of porous
  geomaterials}.
\newblock \bibinfo{type}{Ph.{D.}\ thesis}. University of Glasgow.
\bibitem[{Bisschop and van Mier(2002)}]{BisMie02}
\bibinfo{author}{Bisschop, J.}, \bibinfo{author}{van Mier, J.},
  \bibinfo{year}{2002}.
\newblock \bibinfo{title}{{Effect of aggregates on drying shrinkage
  microcracking in cement-based composites}}.
\newblock \bibinfo{journal}{Materials and Structures} \bibinfo{volume}{35},
  \bibinfo{pages}{453--461}.
\bibitem[{Bolander and Berton(2004)}]{BolBer04}
\bibinfo{author}{Bolander, J.E.}, \bibinfo{author}{Berton, S.},
  \bibinfo{year}{2004}.
\newblock \bibinfo{title}{Simulation of shrinkage induced cracking in cement
  composite overlays}.
\newblock \bibinfo{journal}{Cement and Concrete Composites}
  \bibinfo{volume}{26}, \bibinfo{pages}{861--871}.
\bibitem[{Grassl and Bolander(2016)}]{GraBol16}
\bibinfo{author}{Grassl, P.}, \bibinfo{author}{Bolander, J.},
  \bibinfo{year}{2016}.
\newblock \bibinfo{title}{Three-dimensional network model for coupling of
  fracture and mass transport in quasi-brittle geomaterials}.
\newblock \bibinfo{journal}{Materials} \bibinfo{volume}{9},
  \bibinfo{pages}{782}.
\bibitem[{Grassl and Davies(2011)}]{GraDav11}
\bibinfo{author}{Grassl, P.}, \bibinfo{author}{Davies, T.},
  \bibinfo{year}{2011}.
\newblock \bibinfo{title}{Lattice modelling of corrosion induced cracking and
  bond in reinforced concrete}.
\newblock \bibinfo{journal}{Cement and Concrete Composites}
  \bibinfo{volume}{33}, \bibinfo{pages}{918--924}.
\bibitem[{Grassl and Jir\'{a}sek(2010)}]{GraJir10}
\bibinfo{author}{Grassl, P.}, \bibinfo{author}{Jir\'{a}sek, M.},
  \bibinfo{year}{2010}.
\newblock \bibinfo{title}{Meso-scale approach to modelling the fracture process
  zone of concrete subjected to uniaxial tension}.
\newblock \bibinfo{journal}{International Journal of Solids and Structures}
  \bibinfo{volume}{47}, \bibinfo{pages}{957--968}.
\bibitem[{Grassl et~al.(2010)Grassl, Wong and Buenfeld}]{GraWonBue10}
\bibinfo{author}{Grassl, P.}, \bibinfo{author}{Wong, H.S.},
  \bibinfo{author}{Buenfeld, N.R.}, \bibinfo{year}{2010}.
\newblock \bibinfo{title}{Influence of aggregate size and volume fraction on
  shrinkage induced micro-cracking of concrete and mortar}.
\newblock \bibinfo{journal}{Cement and Concrete Research} \bibinfo{volume}{40},
  \bibinfo{pages}{85--93}.
\bibitem[{Havl{\'a}sek and Jir{\'a}sek(2016)}]{HavJir16}
\bibinfo{author}{Havl{\'a}sek, P.}, \bibinfo{author}{Jir{\'a}sek, M.},
  \bibinfo{year}{2016}.
\newblock \bibinfo{title}{Multiscale modeling of drying shrinkage and creep of
  concrete}.
\newblock \bibinfo{journal}{Cement and Concrete Research} \bibinfo{volume}{85},
  \bibinfo{pages}{55--74}.
\bibitem[{Idiart et~al.(2012)Idiart, Bisschop, Caballero and
  Lura}]{IdiBisCab12}
\bibinfo{author}{Idiart, A.}, \bibinfo{author}{Bisschop, J.},
  \bibinfo{author}{Caballero, A.}, \bibinfo{author}{Lura, P.},
  \bibinfo{year}{2012}.
\newblock \bibinfo{title}{A numerical and experimental study of
  aggregate-induced shrinkage cracking in cementitious composites}.
\newblock \bibinfo{journal}{Cement and Concrete Research} \bibinfo{volume}{42},
  \bibinfo{pages}{272--281}.
\bibitem[{Kanit et~al.(2003)Kanit, Forest, Galliet, Mounoury and
  Jeulin}]{KanForGal03}
\bibinfo{author}{Kanit, T.}, \bibinfo{author}{Forest, S.},
  \bibinfo{author}{Galliet, I.}, \bibinfo{author}{Mounoury, V.},
  \bibinfo{author}{Jeulin, D.}, \bibinfo{year}{2003}.
\newblock \bibinfo{title}{Determination of the size of the representative
  volume element for random composites: statistical and numerical approach}.
\newblock \bibinfo{journal}{International Journal of Solids and Structures}
  \bibinfo{volume}{40}, \bibinfo{pages}{3647--3679}.
\bibitem[{Kawai(1978)}]{Kaw78}
\bibinfo{author}{Kawai, T.}, \bibinfo{year}{1978}.
\newblock \bibinfo{title}{{New discrete models and their application to seismic
  response analysis of structures}}.
\newblock \bibinfo{journal}{Nuclear Engineering and Design}
  \bibinfo{volume}{48}, \bibinfo{pages}{207--229}.
\bibitem[{Lagier et~al.(2011)Lagier, Jourdain, Sa, Benboudjema and
  Colliat}]{LagJouDes11}
\bibinfo{author}{Lagier, F.}, \bibinfo{author}{Jourdain, X.},
  \bibinfo{author}{Sa, C.D.}, \bibinfo{author}{Benboudjema, F.},
  \bibinfo{author}{Colliat, J.B.}, \bibinfo{year}{2011}.
\newblock \bibinfo{title}{Numerical strategies for prediction of drying cracks
  in heterogeneous materials: Comparison upon experimental results}.
\newblock \bibinfo{journal}{Engineering Structures} \bibinfo{volume}{33},
  \bibinfo{pages}{920--931}.
\bibitem[{Lewis et~al.(1996)Lewis, Morgan, Thomas and Seetharamu}]{LewMorTho96}
\bibinfo{author}{Lewis, R.W.}, \bibinfo{author}{Morgan, K.},
  \bibinfo{author}{Thomas, H.R.}, \bibinfo{author}{Seetharamu, K.},
  \bibinfo{year}{1996}.
\newblock \bibinfo{title}{The finite element method in heat transfer analysis}.
\newblock \bibinfo{publisher}{John Wiley \& Sons}.
\bibitem[{Maekawa et~al.(2008)Maekawa, Ishida and Kishi}]{MaeIshKis08}
\bibinfo{author}{Maekawa, K.}, \bibinfo{author}{Ishida, T.},
  \bibinfo{author}{Kishi, T.}, \bibinfo{year}{2008}.
\newblock \bibinfo{title}{Multi-scale modeling of structural concrete}.
\newblock \bibinfo{publisher}{CRC Press}.
\bibitem[{Maruyama and Sasano(2014)}]{MarSas14}
\bibinfo{author}{Maruyama, I.}, \bibinfo{author}{Sasano, H.},
  \bibinfo{year}{2014}.
\newblock \bibinfo{title}{Strain and crack distribution in concrete during
  drying}.
\newblock \bibinfo{journal}{Materials and Structures} \bibinfo{volume}{47},
  \bibinfo{pages}{517--532}.
\bibitem[{Maruyama et~al.(2016)Maruyama, Sasano and Lin}]{MarSasLin16}
\bibinfo{author}{Maruyama, I.}, \bibinfo{author}{Sasano, H.},
  \bibinfo{author}{Lin, M.}, \bibinfo{year}{2016}.
\newblock \bibinfo{title}{Impact of aggregate properties on the development of
  shrinkage-induced cracking in concrete under restraint conditions}.
\newblock \bibinfo{journal}{Cement and Concrete Research} \bibinfo{volume}{85},
  \bibinfo{pages}{82--101}.
\bibitem[{McGuire et~al.(2000)McGuire, Gallagher and Ziemian}]{McgGalZie00}
\bibinfo{author}{McGuire, W.}, \bibinfo{author}{Gallagher, R.H.},
  \bibinfo{author}{Ziemian, R.D.}, \bibinfo{year}{2000}.
\newblock \bibinfo{title}{Matrix structural analysis}.
\newblock \bibinfo{publisher}{J. Wiley}, \bibinfo{address}{New York}.
\bibitem[{Miehe and Koch(2002)}]{MieKoc02}
\bibinfo{author}{Miehe, C.}, \bibinfo{author}{Koch, A.}, \bibinfo{year}{2002}.
\newblock \bibinfo{title}{Computational micro-to-macro transitions of
  discretized microstructures undergoing small strains}.
\newblock \bibinfo{journal}{Journal of Applied Mechanics, ASME}
  \bibinfo{volume}{72}, \bibinfo{pages}{300--317}.
\bibitem[{Nilenius et~al.(2014)Nilenius, Larsson, Lundgren and
  Runesson}]{NilLarLun14}
\bibinfo{author}{Nilenius, F.}, \bibinfo{author}{Larsson, F.},
  \bibinfo{author}{Lundgren, K.}, \bibinfo{author}{Runesson, K.},
  \bibinfo{year}{2014}.
\newblock \bibinfo{title}{Computational homogenization of diffusion in
  three-phase mesoscale concrete}.
\newblock \bibinfo{journal}{Computational Mechanics} \bibinfo{volume}{54},
  \bibinfo{pages}{461--472}.
\bibitem[{Okabe et~al.(2000)Okabe, Boots, Sugihara and Chiu}]{OkaBooSug00}
\bibinfo{author}{Okabe, A.}, \bibinfo{author}{Boots, B.},
  \bibinfo{author}{Sugihara, K.}, \bibinfo{author}{Chiu, S.N.},
  \bibinfo{year}{2000}.
\newblock \bibinfo{title}{Spatial tessellations: Concepts and applications of
  Voronoi diagrams}.
\newblock \bibinfo{publisher}{Wiley New York}.
\bibitem[{Patz\'ak(2012)}]{Pat12}
\bibinfo{author}{Patz\'ak, B.}, \bibinfo{year}{2012}.
\newblock \bibinfo{title}{{OOFEM -- A}n object-oriented simulation tool for
  advanced modeling of materials and structure}.
\newblock \bibinfo{journal}{Acta Polytechnica} \bibinfo{volume}{52},
  \bibinfo{pages}{59--66}.
\bibitem[{Strang(1986)}]{Strang86}
\bibinfo{author}{Strang, G.}, \bibinfo{year}{1986}.
\newblock \bibinfo{title}{Introduction to Applied Mathematics}.
\newblock \bibinfo{publisher}{Wellesley-Cambridge Press},
  \bibinfo{address}{Wellesley, Massachusetts}.
\bibitem[{Witherspoon et~al.(1980)Witherspoon, Wang, Iawai and
  Gale}]{WitWanIwaGal80}
\bibinfo{author}{Witherspoon, P.A.}, \bibinfo{author}{Wang, J.S.Y.},
  \bibinfo{author}{Iawai, K.}, \bibinfo{author}{Gale, J.E.},
  \bibinfo{year}{1980}.
\newblock \bibinfo{title}{Validity of cubic law for fluid flow in a deformable
  rock fracture}.
\newblock \bibinfo{journal}{Water Resour. Res} \bibinfo{volume}{16},
  \bibinfo{pages}{1016--1024}.
\bibitem[{Wong et~al.(2009)Wong, Zobel, Buenfeld and Zimmerman}]{WonZobBue09}
\bibinfo{author}{Wong, H.S.}, \bibinfo{author}{Zobel, M.},
  \bibinfo{author}{Buenfeld, N.R.}, \bibinfo{author}{Zimmerman, R.W.},
  \bibinfo{year}{2009}.
\newblock \bibinfo{title}{Influence of the interfacial transition zone and
  microcracking on the diffusivity, permeability and sorptivity of cement-based
  materials after drying}.
\newblock \bibinfo{journal}{Magazine of Concrete Research}
  \bibinfo{volume}{61}, \bibinfo{pages}{571--589}.
\bibitem[{Wu et~al.(2015)Wu, Wong and Buenfeld}]{WuWonBue15}
\bibinfo{author}{Wu, Z.}, \bibinfo{author}{Wong, H.S.},
  \bibinfo{author}{Buenfeld, N.R.}, \bibinfo{year}{2015}.
\newblock \bibinfo{title}{Influence of drying-induced microcracking and related
  size effects on mass transport properties of concrete}.
\newblock \bibinfo{journal}{Cement and Concrete Research} \bibinfo{volume}{68},
  \bibinfo{pages}{35--48}.

\end{thebibliography}

\appendix
\numberwithin{equation}{section}

\end{document}